\documentclass[aOAps,prb,twocolumn]{revtex4-1}
\usepackage{epsfig}
\usepackage{amsmath}
\usepackage{amsfonts}
\usepackage{braket}
\usepackage{bm}
\usepackage{color}
\usepackage{verbatim} %for multiline-comment
\begin{document}

\title{
Entanglement Chern number for three-dimensional topological insulators:
Characterization by Weyl points of entanglement Hamiltonians
}

\author{Hiromu Araki$^1$, Takahiro Fukui$^2$, and Yasuhiro Hatsugai$^{1,3}$}

\affiliation{$^1$Graduate School of Pure and Applied Sciences, University of Tsukuba, Tsukuba, Ibaraki 305-8571, 
Japan\\
$^2$Department of Physics, Ibaraki University, Mito, Ibaraki 310-8512, Japan\\
$^3$Division of Physics, University of Tsukuba, Tsukuba, Ibaraki 305-8571, Japan\\
}

\date{\today}

\begin{abstract}
We propose characterization of the three-dimensional topological insulator by
using the Chern number for the entanglement Hamiltonian (entanglement Chern number).
Here we take the extensive spin partition of the system, that pulls out the quantum entanglement between
up spin and down spin of the  many-body ground state.
In three dimensions, the topological insulator phase is described by the section entanglement Chern number,
which is the entanglement Chern number for a periodic plane in the Brillouin zone.
The section entanglement Chern number serves as an interpolation of
the $\mathbb{Z}_2$ invariants defined on time-reversal invariant planes.
We find that the change of the section entanglement Chern number protects the Weyl point of the entanglement Hamiltonian
and the parity of the number of Weyl points distinguishes the strong topological insulator phase
from the weak topological insulator phase.
\end{abstract}

\maketitle
%\tableofcontents

\section{Introduction}

Classification of matter from a topological point of view is nowadays one of fundamental ideas for
understanding various phases of matter. \cite{Hasan:2010fk,Qi:2011kx}
Topological phases for  noninteracting fermion systems have been classified under some
fundamental symmetries.
\cite{PhysRevB.78.195125,PhysRevB.78.195424,DOI:10.1063/1.3149495,1367-2630-12-6-065010}
Over a decade, interacting cases have been studied, revealing that interactions
enrich the topological phases.
\cite{
PhysRevLett.100.156401,
10.1038/nphys1606,
PhysRevLett.104.106408,
PhysRevB.81.134509,
PhysRevLett.105.256803,
PhysRevB.83.075103,
PhysRevB.83.075102,
PhysRevB.85.045130,
PhysRevB.86.125119, 
PhysRevB.93.115131,
PhysRevLett.118.147001}
A topologically nontrivial system shows gapless edge states even though the bulk is insulating,
\cite{Hatsugai:1993fk}
which may open the possibility of new devices.
\cite{Konig766, PhysRevB.84.233101, Mourik1003}

The topological classification needs suitable topological invariants (topological numbers)
to distinguish the phases.
Therefore, it is important to establish the practical method of computing topological numbers.
\cite{doi:10.1143/JPSJ.74.1674,PhysRevB.85.165126, PhysRevX.2.031008}
Quantum Hall states are characterized by $\mathbb{Z}$ as the Chern number and topological insulators
with time-reversal symmetry are by $\mathbb{Z}_2$ numbers.
\cite{PhysRevLett.95.146802,PhysRevLett.95.226801,PhysRevB.74.195312}
In two dimensions,
 a single $\mathbb{Z}_2$ number is enough to characterize the topological insulator.
This topological number
has been proposed for the Kane--Mele model of the quantum spin Hall effect,
\cite{PhysRevLett.95.146802,PhysRevLett.95.226801}
and  established as an adiabatic time-reversal polarization  for the spin pump model.
\cite{PhysRevB.74.195312}
Alternatively, based on an effective field theory, the idea of understanding the
$\mathbb{Z}_2$ invariant as the parity of Chern numbers has been presented.
\cite{PhysRevB.78.195424}
Based on these studies, several practical methods of computing the $\mathbb{Z}_2$ invariant
for generic systems have been proposed.
\cite{Fukui:2007kq,Soluyanov:2011aa}

The quantum entanglement is an essential concept for quantum mechanics and the entanglement
entropy is also used as a fundamental fingerprint to characterize quantum phase transitions.
\cite{PhysRevLett.90.227902,PhysRevB.73.245115}
%It has also been used to characterize the fractional quantum Hall states.
%\cite{PhysRevLett.101.010504}
Its topological term (topological entanglement entropy) is one of the signatures of the fractional quantum Hall states, \cite{PhysRevLett.96.110404,PhysRevLett.96.110405,PhysRevLett.98.060401,PhysRevB.76.125310,PhysRevB.78.035320}
and the entanglement spectrum has sufficient information to identify them.
\cite{PhysRevLett.101.010504}
To investigate the topological insulator phase by using the quantum entanglement between subsystems,
not only simple spatial partition but the extensive partition are useful.
\cite{PhysRevB.84.195103,PhysRevLett.113.106801}
The entanglement Chern number is the topological number focusing on the quantum entanglement
in a many-body ground state.
\cite{doi:10.7566/JPSJ.83.113705}
When the Kane--Mele model is regarded as
two copies of Haldane model \cite{PhysRevLett.61.2015}
with nontrivial Chern numbers,
it is useful to characterize the quantum spin Hall state using Chern numbers for these subsystems.
The spin Chern number is one of them,
\cite{PhysRevLett.97.036808,Fukui:2007sf}
and the entanglement Chern number is another one
which is the
Chern number for the entanglement Hamiltonian.
We have demonstrated that two-dimensional $\mathbb{Z}_2$
topological insulators are characterized by the entanglement Chern number.\cite{doi:10.7566/JPSJ.84.043703,
doi:10.7566/JPSJ.85.043706}
%The Fu--Kane--Mele model, a tight-binding model of $s$ states on the diamond lattice with a spin-orbit interaction, is the generalization of the quantum spin Hall model to three dimensions. 
A generalization of the Kane--Mele model to three dimensions is spin-orbit coupled 
tight-binding model on a zinc-blende lattice. The 
sublattice potential breaks inversion symmetry, then the system becomes a Weyl semimetal, 
\cite{PhysRevLett.108.140405, PhysRevB.87.245112} In the case that the sublattices are equivalent, a 
spatial anisotropy gaps out the spectrum and the topological insulator phase appears. 
\cite{PhysRevLett.98.106803}
In three dimensions,  it is known that the topological phases are classified by four $\mathbb{Z}_2$ numbers and there are
two distinguished topological phases:
the strong topological insulator (STI) phase and the weak topological insulator (WTI) phase.\cite{PhysRevB.79.195322,PhysRevB.75.121306,PhysRevLett.98.106803}
The STI is characterized by a three-dimensional topological index, and has an odd number of surface Dirac cones on every surface.
In contrast, the WTI is adiabatically connected to stacked layers of two-dimensional topological insulators, and has an even number of Dirac cones on side surfaces.
The purpose of this paper is to characterize these topological phases by the entanglement Chern number.

In this paper, we discuss the characterization of the WTI and the STI in view of the entanglement Chern number and the entanglement
Hamiltonians.
In three dimensions, the $\mathbb{Z}_2$ invariants can be defined for six time-reversal-invariant planes.
In contrast, the entanglement Chern number is defined for any section, implying that the section entanglement Chern number interpolates the $\mathbb{Z}_2$ invariants
continuously. It follows that the entanglement Hamiltonians should have some gapless points along such an
interpolation when the entanglement Chern number changes.
Since the parities of  these gapless points are topologically protected and the entanglement Hamiltonians
have ``broken'' time-reversal symmetry, we refer to the gapless points of the entanglement Hamiltonian as the Weyl points.
We demonstrate that the entanglement Chern number and the entanglement Hamiltonians simply characterize the topological insulator in three dimensions.
This paper is organized as follows.
In Secs. \ref{s:z2} and \ref{s:ec}, we give a brief review of the $\mathbb{Z}_2$ invariant and the entanglement Chern number, respectively.
In Secs. \ref{s:wd} and \ref{s:fkm}, we examine the STI and WTI phases of
the Fu--Kane--Mele  model and the Wilson--Dirac model, respectively,
in terms of the entanglement Chern number.
We give concluding remarks in Sec. \ref{s:cl}.

\section{$\mathbb{Z}_2$ invariant for topological insulators}
\label{s:z2}%-------

\begin{figure}[t]
 \centering
 \epsfig{file=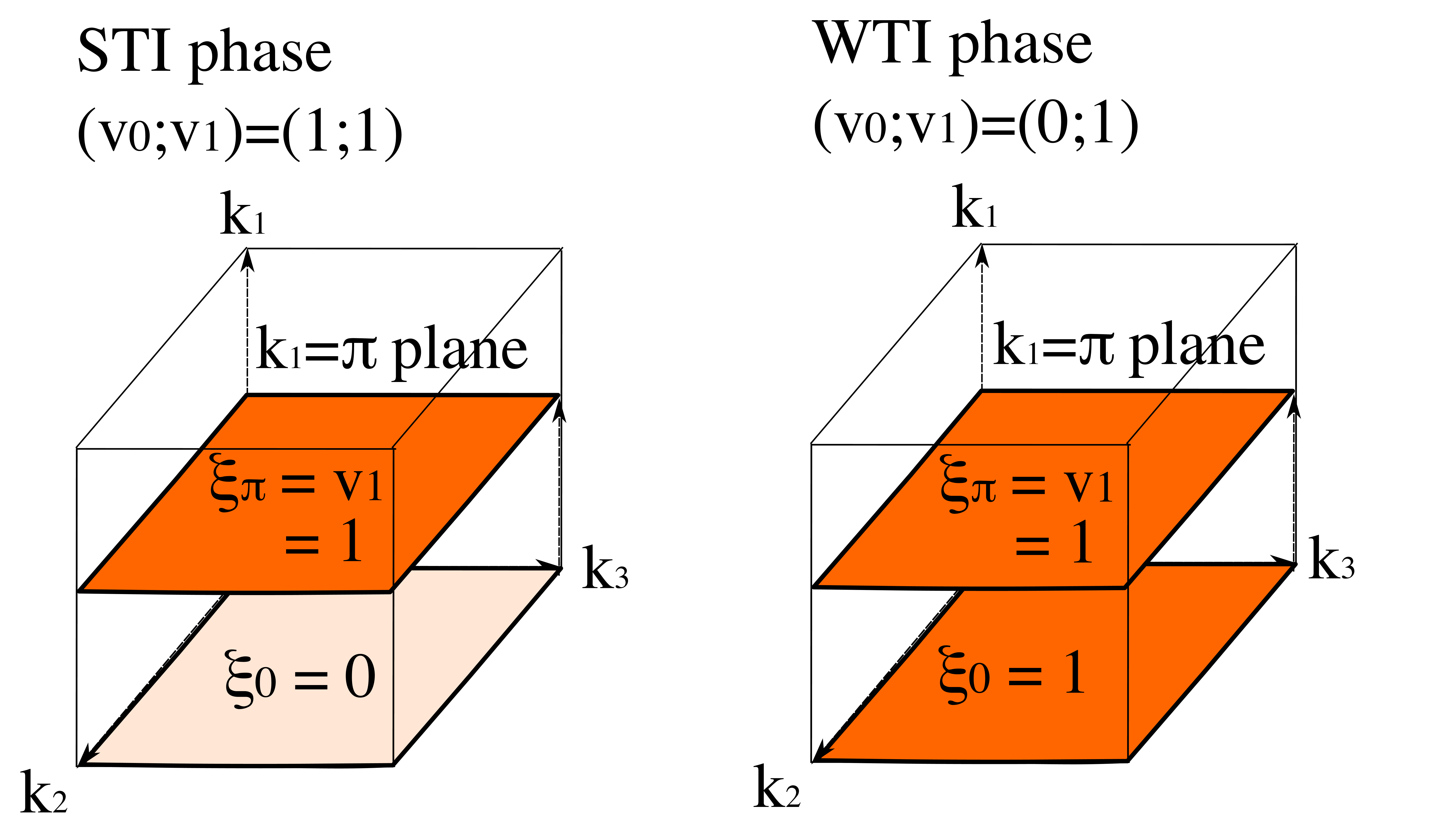,width=0.48\textwidth}
 \caption{(Color online) Schematic picture of $\mathbb{Z}_2$ indices defined for time-reversal-invariant planes.
 The Brillouin zone of the STI with $(\nu_0;\nu_1)=(1;1)$ (left) and the WTI with (0;1) (right) are shown.
 $\bm{b}_1$, $\bm{b}_2$ and $\bm{b}_3$ are the reciprocal lattice vectors.
 The colored regions are the time-reversal-invariant planes. The weak $\mathbb{Z}_2$ numbers $\nu_i ~(i=1, 2, 3)$ are defined for each of the time-reversal-invariant planes.
 For the STI (WTI) phases, the entanglement Hamiltonian has odd (even) pairs of Weyl points between time-reversal-invariant planes.
 }
 \label{fig:Z2indices}
\end{figure}

Now we briefly introduce the $\mathbb{Z}_2$ invariants.
Let $\Lambda_i$ ($i=1,2,...,8$) be the time-reversal-invariant momentum.
The topological indices $\delta(\Lambda_i)$ are defined at time-reversal-invariant momentum as
\begin{equation}
 \delta (\Lambda_i) = \frac{\mathrm{Pf}[w(\Lambda_i)]}{\sqrt{\mathrm{det}[w(\Lambda_i)]}}.
\end{equation}
Here $w_{\alpha\beta}(\bm{k}) = \braket{u_{\alpha -\bm{\bm k}}|\Theta|u_{\beta \bm{k}}}$
is defined for the Bloch state $\ket{u_{\alpha \bm{k}}}$ and
 $\Theta$ is the time-reversal symmetry operator.
Let $\bm b_j$ $(i=1,2,3)$ be the normalized reciprocal lattice vector (reciprocal vector of length 1).
Then, the wave vector is spanned by\
\begin{equation}
 \label{eq:def of params in BZ}
 \bm k=k_1\bm b_1+k_2\bm b_2+k_3 \bm b_3 \mbox{ ($0\leq k_1,k_2,k_3<2\pi$)}.
\end{equation}
The topological phases in three dimensions are characterized by four $\mathbb{Z}_2$ indices $\nu_0(\nu_1\nu_2\nu_3)$
\cite{PhysRevLett.95.146802,PhysRevB.75.121306,PhysRevLett.98.106803}.
The strong $\mathbb{Z}_2$ number $\nu_0$ is defined as
\begin{equation}
 (-1)^{\nu_0} = \prod_{i=1}^8 \delta (\Lambda_i).
\end{equation}
On the other hand, the weak $\mathbb{Z}_2$ indices $\nu_n~(n=1,2,3)$
are defined on the time-reversal-invariant plane $P_n$, with fixed $k_n=\pi$, as
\begin{equation}
  (-1)^{\nu_n} = \prod_{\Lambda_i \in {P_n}}  \delta (\Lambda_i).
\end{equation}

Alternatively, such $\mathbb{Z}_2$ invariants can be viewed as {\it the section $\mathbb{Z}_2$ invariants},
\cite{PhysRevB.79.195322,PhysRevB.75.121306,PhysRevLett.98.106803} which
we briefly review to fix our notations in the following sections.
Let us set one component of momentum, say, $k_1$,
as the time-reversal-invariant momentum, $k_1=0$ or $\pi$.
Then, such a Hamiltonian as is defined on the time-reversal-invariant plane $k_1=0$ or $\pi$
corresponds to a topological insulator in two dimensions, which is, as a result,
described by the single $\mathbb{Z}_2$ invariant.
Let $\xi_0$ and $\xi_{\pi}$ be the $\mathbb{Z}_2$ invariants on the $k_1=0$ and $k_1=\pi$
time-reversal-invariant plane, respectively.
These two $\mathbb{Z}_2$ invariants may be called {\it section $\mathbb{Z}_2$ numbers} on the time-reversal-invariant planes
along the $\bm b_1$ axis.
Likewise, we can define four other section $\mathbb{Z}_2$ invariants, $\eta_{0,\pi}$ and $\zeta_{0,\pi}$
on the time-reversal-invariant planes, $k_2=0, \pi$ and $k_3=0, \pi$, respectively.
Since there are two constraints $\xi_0+\xi_{\pi}=\eta_0+\eta_{\pi}=\zeta_0+\zeta_{\pi}$ (mod 2)  among six invariants,
we have four independent $\mathbb{Z}_2$ invariants, which are conventionally defined by
$\nu_0(\nu_1\nu_2\nu_3)$ with $\nu_0=\xi_0+\xi_{\pi}$ (mod 2), $\nu_1=\xi_{\pi}$, $\nu_2=\eta_{\pi}$, and
$\nu_3=\zeta_{\pi}$.
The STI specified by $\nu_0=1$
has the different $\mathbb{Z}_2$ indices between parallel time-reversal-invariant planes, whereas
the WTI specified $\nu_0=0$ has the same but nontrivial $\mathbb{Z}_2$ indices
between them, as described in Fig. \ref{fig:Z2indices}.
Such a classification is also possible by the entanglement Chern number
\cite{doi:10.7566/JPSJ.83.113705,doi:10.7566/JPSJ.84.043703,doi:10.7566/JPSJ.85.043706}
computed on the time-reversal-invariant planes, as will be reviewed in the following section.

\section{Entanglement Chern numbers}
\label{s:ec}%-------

\subsection{Entanglement Hamiltonian}

Let $|\Psi\rangle$ be the many-body ground state of a noninteracting system.
We divide the system into two parts, say, A and B, and
consider quantum entanglement between them in $|\Psi\rangle$.
To this end, let us introduce the density matrix $\rho = \ket{\Psi}\bra{\Psi}$.
Tracing out B in $\rho$,
we obtain the reduced density matrix for the subsystem A such that
$\rho_{\rm A} = \mathrm{Tr_B}\rho=\mathrm{Tr_B}\ket{\Psi}\bra{\Psi}$.
As shown in Ref.~\onlinecite{0305-4470-36-14-101}, the reduced density matrix is also given as
\begin{equation}
 \rho_{\rm A} \propto e^{-\mathcal{H}_{\rm A}},
\end{equation}
where the $\mathcal{H}_{\rm A}$ is the entanglement Hamiltonian of the subsystem A.

\subsection{Entanglement Chern numbers and $\mathbb{Z}_2$ invariants}
Let us first introduce the entanglement Chern number in two dimensions.
Let ${\cal B}$ be the two-dimensional Brillouin zone spanned by the wave vector $\bm k = (k_1,k_2)$.
Then, the entanglement Chern number is defined as the Chern number
for the entanglement Hamiltonian $\mathcal{H}_{\rm A}$:
\begin{equation}
 c_{\rm A} = \frac{1}{2\pi i}\int_{\cal B} F_{12}(\bm{k})d^2k ,
\end{equation}
where $F_{12}(\bm{k})=\epsilon_{ij}\partial_{k_i}\psi_{\rm A}^\dagger (\bm k)\partial_{k_j}\psi_{\rm A}(\bm k)$
is the Berry curvature
of the ground state (negative energy) multiplet $\psi_{\rm A}$ of the entanglement Hamiltonian  $\mathcal{H}_{\rm A}$.
For free fermion systems, it is convenient to use the correlation matrix instead
of the entanglement Hamiltonians. \cite{0305-4470-36-14-101, doi:10.7566/JPSJ.83.113705,
doi:10.7566/JPSJ.85.043706}
The correlation matrix for the subsystem A is defined as
$(C_{\rm A})_{ij} = \braket{c_i^\dagger c_j} = \mathrm{Tr}[\rho_{\rm A} c_i^\dagger c_j]$, where
$c_i^\dagger$ is the creation operator of an electron in the subsystem A.
Then, it turns out \cite{0305-4470-36-14-101} that $C_{\rm A}$ is related with ${\cal H}_{\rm A}$ such that
\begin{equation}
 \mathcal{H}_{\rm A}^T = \ln [(1-C_{\rm A})/C_{\rm A}].
\end{equation}
By calculating the Chern number for the correlation matrix, we can obtain the Chern number for the entanglement Hamiltonian.
\cite{doi:10.7566/JPSJ.83.113705}

The entanglement Chern number was successfully applied to Kane--Mele model choosing the up and down spins perpendicular the surface
for the partition A and B.
\cite{doi:10.7566/JPSJ.83.113705,doi:10.7566/JPSJ.85.043706}
Let $c_{\rm \uparrow}$ and $c_{\rm \downarrow}$ be the entanglement Chern number for up- and down-spin, respectively.
There, the relationship between the $\mathbb{Z}_2$ invariant and entanglement Chern number is as follows:
\begin{alignat}1
&\nu=0 \leftrightarrow \mbox{even } c_{\rm \uparrow},
\nonumber\\
&\nu=1 \leftrightarrow \mbox{odd } c_{\rm \uparrow}.
\label{Z2ecCor}%-------
\end{alignat}
With the time-reversal symmetry, $c_{\rm \downarrow}$ equals $-c_{\rm \uparrow}$.
The entanglement Chern number with respect to spin plays the role of the spin Chern number.
\cite{PhysRevLett.97.036808,Fukui:2007sf}

\subsection{Section entanglement Chern numbers and Weyl points of entanglement Hamiltonians}
\label{s:ECN and Weyl}

The {\it section} entanglement Chern number in three dimensions is defined as the entanglement Chern number for a two-dimensional periodic
plane in the three-dimensional Brillouin zone.
Namely, in Fig. \ref{fig:Z2indices}, one can calculate the entanglement Chern number for any sections parallel to the time-reversal-invariant planes.
This implies that
the section $\mathbb{Z}_2$ invariants on the time-reversal-invariant planes can be interpolated by
the {\it section} entanglement Chern number.
For a STI, the entanglement Chern numbers on the two time-reversal-invariant planes, as well as the $\mathbb{Z}_2$ invariants, are different
between these time-reversal-invariant planes.
Therefore, if we compute the section entanglement Chern number continuously along, e.g., the $\bm b_1$ axis,
some gapless points of the entanglement Hamiltonian accompany the change of the entanglement Chern number,
implying that the STI phase can be characterized by these gapless points of the entanglement Hamiltonian.
Note that the section Chern number for the original Hamiltonian is always trivial because of the time-reversal symmetry.

The entanglement Hamiltonians retain the spatial inversion symmetry,
but have ``broken'' time-reversal symmetry (i.e., they do not support time-reversal symmetry),
therefore one pair of gapless points emerges from the time-reversal-invariant momentum.
These gapless points have opposite chirality, $+$ and $-$.
This scenario has the same topological origin as the emergence of the gapless points in a Weyl semimetal,
\cite{Murakami:2007uq,Burkov:2011ab}
so we refer to the gapless points of the entanglement Hamiltonian as the Weyl points.
Because of the linear dispersion, a Weyl point of the entanglement Hamiltonians changes the section entanglement Chern number by $\pm 1$.

Now we define $N_W^{(i)}$ as the number of Weyl points between the $k_i=0$ time-reversal-invariant plane
and the $k_i=\pi$ time-reversal-invariant plane in the Brillouin zone.
For the WTI phase, the $\mathbb{Z}_2$  indices between the time-reversal-invariant planes are
the same modulo 2 and nontrivial at least in a certain direction.
Correspondingly, the section entanglement Chern number for the time-reversal-invariant planes are the same modulo 2.
On the other hand, for the STI phase, the $\mathbb{Z}_2$  indices between the time-reversal-invariant planes are
different.
Correspondingly, the section entanglement Chern number for the time-reversal-invariant planes are also different.
Thus, the STI (WTI) phase has odd (even) number of $N_W^{(i)}$ for any $i$:
\begin{alignat}1
&\mbox{WTI phase} \rightarrow \mbox{even $N_W^{(i)}$},
\nonumber\\
&\mbox{STI phase} \rightarrow \mbox{odd $N_W^{(i)}$}.
\end{alignat}
In the following sections, we demonstrate it for two examples:
the Fu--Kane--Mele model and the Wilson--Dirac model.

In passing, we add a comment that  another merit  of using the entanglement Chern number is that
the section entanglement Chern number is potentially applicable to topological insulators with
interactions and/or broken time-reversal-invariant perturbations.

\section{Fu--Kane--Mele model}
\label{s:fkm}%-------

We first examine the Fu--Kane--Mele model \cite{PhysRevLett.98.106803}:
\begin{equation}
 \mathcal{H} = \sum_{\braket{ij}}t_{ij}c_i^\dagger c_j + i \lambda_{\rm SO} \sum_{\braket{\braket{ij}}}
c_i^\dagger\bm{s}\cdot\left(\bm{d}_{ij}^1\times\bm{d}_{ij}^2\right)c_j.
\end{equation}
Here, $c_i^\mathrm{T} = (c_{i,\uparrow}, c_{i, \downarrow})$ is creation operators of an electron at site $i$
and $\bm{s}$ is a spin operator of the electron.
The first and the second term of the above Hamiltonian are 
a nearest-neighbor hopping term and a next-nearest-neighbor
spin-orbit interaction term, respectively.
$\bm{d}_{ij}^{1,2}$ are the two nearest-neighbor bonds traversing from the site {\it i} to
the site {\it j}.

The momentum representation of the Hamiltonian using the basis $c_{\bm{k}}^\mathrm{T}=
(c_{\bm{k}A\uparrow}, c_{\bm{k}A\downarrow}, c_{\bm{k}B\uparrow}, c_{\bm{k}B\downarrow})$ is given by
\begin{equation}
 H_{\rm FKM}(\bm{k}) =
  \left(
   \begin{array}{cc}
    P(\bm{k}) & S(\bm{k}) \\
    S(\bm{k})^* & -P(\bm{k}) \\
   \end{array}
  \right),
\end{equation}
where the 2 $\times$ 2 matrices $S(\bm{k})$ and $P(\bm{k})$ are
\begin{alignat}1
& S(\bm{k}) = \sum_{\mu}t_\mu e^{i\bm{k}\cdot \bm{d}_\mu} \\
& P(\bm{k}) = i \lambda_{\rm SO} \sum_{\mu, \nu} e^{i\bm{k}\cdot (-\bm{d}_\mu+\bm{d}_\nu)}
  \bm{s}\cdot(-\bm{d}_\mu\times\bm{d}_\nu).
\end{alignat}
Here,
$\bm{d}_\mu$ ($\mu = 1,2,3,4$)
stands for the bond vector associate with the $\mu$th nearest neighbor and
$t_\mu$ is the nearest-neighbor-hopping constant with bond direction $\bm{d}_\mu$,
The Hamiltonian $H_{\rm FKM}$ respects the time-reversal symmetry and inversion symmetry.
For the isotropic case $t_\mu = t$ $(\mu=1,2,3,4)$,
the system has three Dirac cones at time-reversal-invariant momenta,
$X_1=(\bm{b}_2+\bm{b}_3)/2$, $X_2=(\bm{b}_3+\bm{b}_1)/2$ and $X_3=(\bm{b}_1+\bm{b}_2)/2$.
For the anisotropic case, the system can be a topological insulator.
Table \ref{t:FKM} shows topological phases associated with the anisotropy about $t_1$- and $t_2$-bond directions.
\cite{PhysRevB.76.045302}

\begin{table}[hhb]
 \begin{tabular}{c||c|c|c} \hline
  $t_1/t$ & $-3<t_1/t<-1$ & $-1<t_1/t<1$ & $1<t_1/t<3$ \\ \hline
  $\nu_0(\nu_1\nu_2\nu_3)$ & 1(000) & 0(111) & 1(111) \\ \hline
 \end{tabular}
 \begin{tabular}{c||c|c|c} \hline
  $t_2/t$ & $-3<t_2/t<-1$ & $-1<t_2/t<1$ & $1<t_2/t<3$\\ \hline
  $\nu_0(\nu_1\nu_2\nu_3)$ & 1(000) & 0(100) & 1(100)  \\ \hline
 \end{tabular}
 \caption{(Color online) The top (bottom) table shows nontrivial topological phases of the Fu--Kane--Mele model
 with modification of the hopping parameter $t_1$ ($t_2$). The other hopping parameters
 are fixed to $t$.
 $t_i/t<-3$ and $3< t_i/t$ ($i=1,2$) belong to an ordinary insulating phase.}
\label{t:FKM}%-------
\end{table}

Now we investigate $(\nu_1 \nu_2 \nu_3)=(111)$ and $(100)$ phases in Table \ref{t:FKM}.
To obtain the entanglement Hamiltonians, we take the same partition with respect to the spin as that in Ref.~\onlinecite{doi:10.7566/JPSJ.85.043706}.
However, in three dimensions, the choice of the spin direction is more ambiguous than that in two dimensions.
In the following, we exemplify two kinds of the section entanglement Chern number with different choices of spin axis.
The first one is the $t_1$-bond direction and the second one is the $t_2$-bond direction.
This means that we focus on the spin along these directions and consider the purified (decoupled) limit of
spins.\cite{doi:10.7566/JPSJ.84.043703}
The system is decomposed into up spin and down spin by the entanglement Hamiltonians as follows.
Let $\bm n$ be a $\bm n$-directional vector in real space: then $\bm{n}\cdot\bm{s}$ is
a spin operator along the $\bm n$ direction.
Let $U_{\bm{n}}$ be a $SU(2)$ matrix that diagonalizes the $\bm{n}\cdot\bm{s}$: then we rotate
the original Hamiltonian as
\begin{equation}
 H(\bm{k})\rightarrow U_{\bm{n}} H(\bm{k}) U_{\bm{n}}^\dagger
\end{equation}
to take a new representation by the quantized spin basis along the $\bm n$ direction.
We can easily calculate the correlation matrix with respect to the spins along the $\bm{n}$ direction:
then we get the entanglement Hamiltonians $H_\uparrow(\bm{k})$ and $H_\downarrow(\bm{k})$ for the spins.
In the following, we focus on the entanglement Chern number for $H_\uparrow(\bm{k})$.

\subsection{STI phase}

\begin{figure}[th]
 \begin{center}
  \epsfig{file=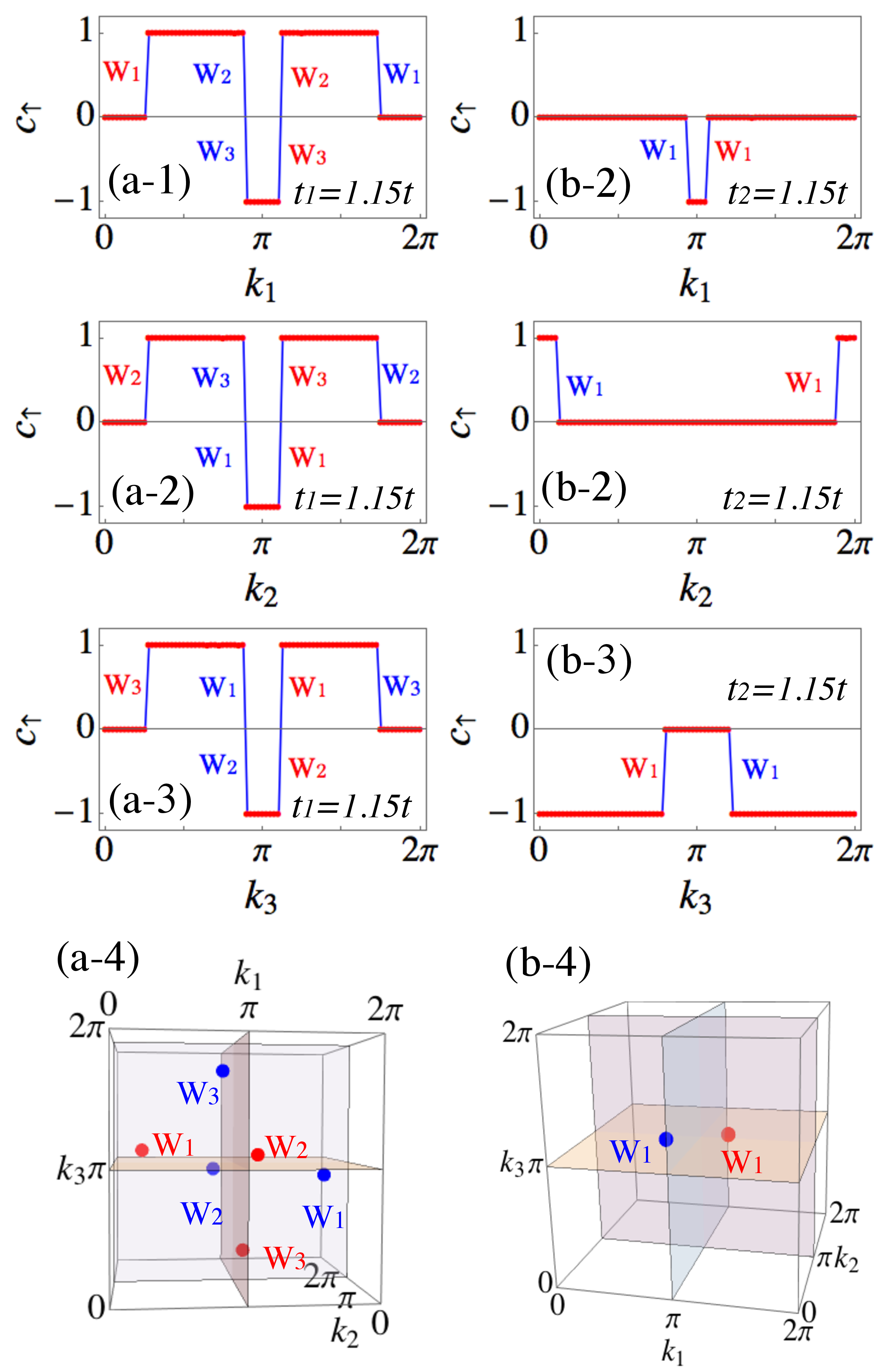,width=0.48\textwidth}
 \end{center}
 \caption{(Color online)
 The section entanglement Chern number and the Weyl points of $H_\uparrow(\bm{k})$ for the STI phases.
 (a) Hopping parameters are  $(t_1,t_2,t_3,t_4)=(1.15t,t,t,t)$ and
 the topological phase is 1(111).
 To obtain the entanglement Hamiltonians, we set $\bm{n}$ as the $t_1$-bond direction.
 (b) Hopping parameters are  $(t_1,t_2,t_3,t_4)=(t,1.15t,t,t)$ and
 the topological phase is 1(100).
 In this case, we set $\bm{n}$ as the $t_2$-bond direction.
 In panels (1)-(3) in both cases, the section entanglement Chern number are shown
 as a function of fixed $k_i$ defined in Eq. (\ref{eq:def of params in BZ}).
 Panels (a-4) and (b-4) show the Weyl points of $H_\uparrow({\bm k})$ in the Brillouin zone.
 %\cite{Supp}
 A red (blue) point represents the Weyl point of chirality $+$ ($-$).
 Panels (1)-(3) show that the section entanglement Chern numbers interpolate the $\mathbb{Z}_2$ topological indices.
 Corresponding to the change of the section entanglement Chern number, the Weyl point of appropriate chirality appears.
 $N_W^{(i)}=3~(i=1,2,3)$ and $N_W^{(i)}=1$ are odd in (a) and (b), respectively.
 For both cases, we set $\lambda_{\rm SO}=0.125t$.
 }
 \label{fig:FKM STI}
\end{figure}

Figure \ref{fig:FKM STI} shows the section entanglement Chern number for the STI phase of
(a) $1(111)$ and (b) $1(100)$.
$N_W^{(i)}=3~(i=1,2,3)$ in (a) and $N_W^{(i)}=1$ in (b).
In both cases, the parity of $N_W^{(i)}$ is odd, which is consistent with the characterization of STI explained
in Sec. \ref{s:ECN and Weyl}.
In addition, Eq. (\ref{Z2ecCor}) tells us that $\xi_0=0$ and $\xi_{\pi}=1$, showing $\nu_0=1$.
The Weyl pairs emerge from $X_1, X_2$ and $X_3$.
By increasing $t_1$ they finally annihilate at $L=(\bm{b}_1+\bm{b}_2+\bm{b}_3)/2$ when $t_1=3$.
On the other hand, by increasing $t_2$ in the (b) case, the Weyl pair emerges only from $X_1$ and annihilates
at $L_1 = \bm{b}_1/2$.
The section entanglement Chern number is symmetric about the $k_i=\pi$ ($i=1,2,3$) point because of the inversion symmetry.
In addition, there empirically holds the sum rule that the total entanglement Chern number equals the Chern number for
the original Hamiltonian.
For the topological insulator phase, the Chern number is zero because of the time-reversal symmetry.
The relation holds for each section entanglement Chern number:
$c_{\rm \uparrow} + c_{\rm \downarrow} = 0$.

\subsection{WTI phase}

\begin{figure}[th]
 \begin{center}
  \epsfig{file=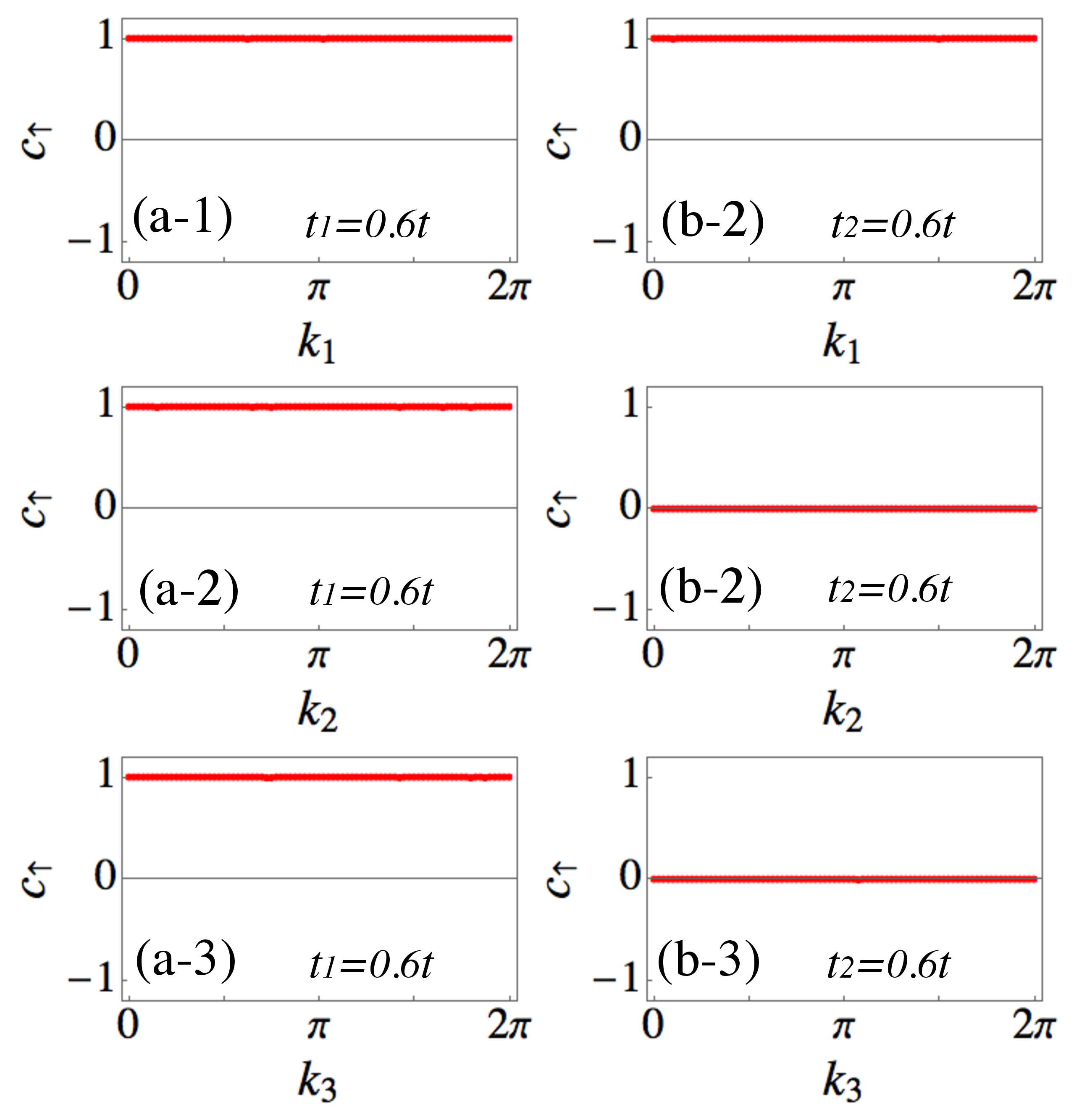,width=0.48\textwidth}
 \end{center}
 \caption{(Color online)
 The section entanglement Chern number for $H_\uparrow(\bm{k})$ for the WTI phases.
 (a) Hopping parameters are  $(t_1,t_2,t_3,t_4)=(0.6t,t,t,t)$ and
 the topological phase is 0(111).
 To obtain the entanglement Hamiltonians, we set $\bm{n}$ as the $t_1$-bond direction.
 (b) Hopping parameters are  $(t_1,t_2,t_3,t_4)=(t,0.6t,t,t)$ and
 the topological phase is 0(100).
 In this case, we set $\bm{n}$ as the $t_2$-bond direction.
 In panels (1)-(3), the section entanglement Chern number are shown
 as a function of fixed $k_i$ defined in Eq. (\ref{eq:def of params in BZ}).
 Panels (1)-(3) show that the section entanglement Chern numbers correspond to the $\mathbb{Z}_2$ topological indices.
 In both cased, there are no Weyl points and the section entanglement Chern number does not change.
 For both cases, we set $\lambda_{\rm SO}=0.125t$.
 }
 \label{fig:FKM WTI}
\end{figure}

Figure \ref{fig:FKM WTI} shows the section entanglement Chern number for the WTI phase with the $\mathbb{Z}_2$ indices
(a) $0(111)$ and (b) $0(100)$.
The quantized spin axis is in the (a) $t_1$-bond direction and (b) $t_2$-bond direction.
In (a),  $\xi_0=1$ and $\xi_{\pi}=1$, showing $\nu_0=0$.
In this case, the entanglement spectrum is fully gapped and has no Weyl point.
Even $N_W^{(i)}~(i=1,2,3)$ corresponds to the WTI phases.
Taking the $t_1\to 0$ limit, the system is adiabatically connected to the stacked Kane--Mele model\cite{PhysRevLett.95.146802,PhysRevLett.95.226801}
along with the $t_1$-bond direction.
Hence the section entanglement Chern number equals that of the Kane--Mele model\cite{doi:10.7566/JPSJ.85.043706}.
In (b), the $t_2 \rightarrow 0$ limit represents the stacked Kane--Mele model along the $t_2$-bond direction.
It is also found that the $\xi_i$, $\eta_i$ and $\zeta_i$ ($i=0, \pi$) are the same as the $\mathbb{Z}_2$ indices,
because of the correspondence in Eq. (\ref{Z2ecCor}).

\section{Wilson--Dirac  model}
\label{s:wd}%-------

We next examine the Wilson--Dirac model, which is one of the typical and simple models for
topological insulators. It is defined on a cubic lattice,
whose Hamiltonian in the momentum representation is given by
\begin{alignat}1
H_{\rm WD}=t_0\sum_j \sin k_j \gamma_j+\left[m+b\sum_j(\cos k_j-1)\right]\gamma_4,
\end{alignat}
where the $\gamma$-matrices are defined by
$\gamma_j=\tau_3\sigma_j$ for $j=1,2,3$ and $\gamma_4=\tau_1$ where
$\sigma_i$ and $\tau_i$ are the Pauli matrices associated with spin degree of
freedom and pseudospin degree of freedom, respectively.
The topological phases are summarized in Table \ref{t:PD}.

\begin{table}[th]
\begin{center}
\begin{tabular}{c|c|c|c}
\hline
$m/b$&$0<m/b<2$&$2<m/b<4$&$4<m/b<6$\\
\hline\hline
$\nu_0(\nu_1\nu_2\nu_3)$&1(000)&0(111)&1(111)\\
\hline
\end{tabular}
\caption{(Color online)
 Nontrivial topological phases of the Wilson--Dirac model.
 $m/b<0$ and $6<m/b$ belong to an ordinary insulating phase.
}
\label{t:PD}%-----
\end{center}
\end{table}

To calculate the entanglement Hamiltonians, we need to choose a suitable partition of the system.
The partition is required to give the  gapped entanglement spectrum at {\it any} time-reversal-invariant planes.
Otherwise, there is no correspondence between the $\mathbb{Z}_2$ invariants and the entanglement Chern number.
Numerical calculation shows that a simple partition with respect to
$\ket{\uparrow}_\sigma$ or $\ket{\uparrow}_\tau$, which are one-particle state of up spin and up pseudospin, respectively,
does not satisfy the condition: at some time-reversal-invariant planes, the spectrum is gapped, but at other time-reversal-invariant
planes, the spectrum is gapless.
To avoid this difficulty, we introduce a transformation matrix $M=\bm n\cdot \bm\gamma$, where
\begin{alignat}1
%&M=
%\nonumber\\
&\bm n=(\sin\theta_1\sin\theta_2\sin\theta_3,\,
\sin\theta_1\sin\theta_2\cos\theta_3, \nonumber\\
&~~~~~~\ \sin\theta_1\cos\theta_2,\,\cos\theta_1),
\nonumber\\
&\bm \gamma=(\gamma_1,\gamma_2,\gamma_3,\gamma_4),
\end{alignat}
and make a rotation for the $\gamma$ matrices such that $\bm\gamma\rightarrow M\bm\gamma M$.
In what follow, we use $\theta_1=\theta_2=\theta_3=\pi/3$ for numerical calculations,
and the partition is the up-spin part of $\tau_3$ defined by the projection
$P=(1+\tau_3)/2$ for the rotated matrices.
If we choose any other parameters $\theta_j$, the result is the same,
as long as the entanglement spectrum is gapped at any time-reversal-invariant planes.

\begin{figure}[th]
\begin{center}
 \epsfig{file=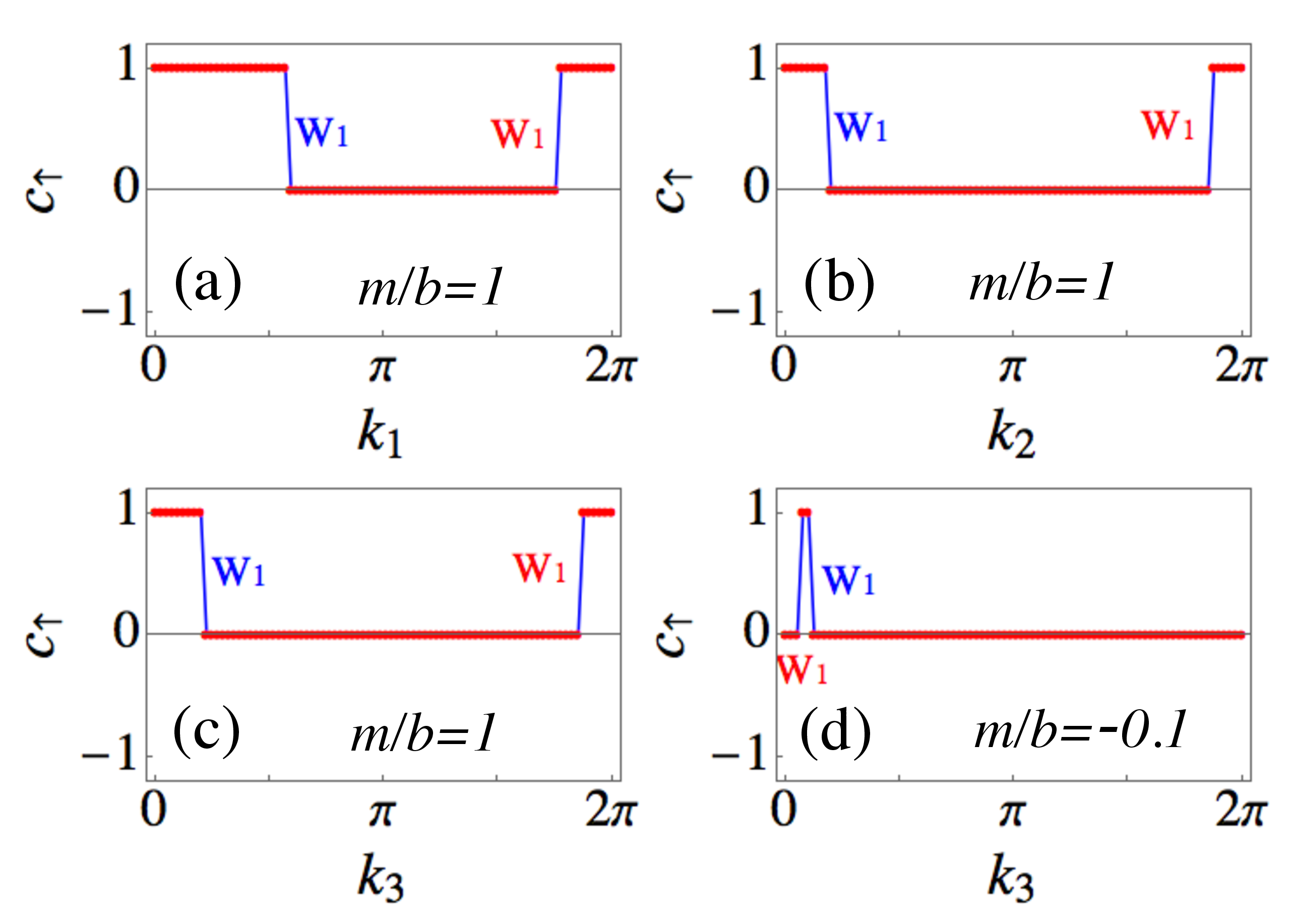,width=0.48\textwidth}
\caption{(Color online)
Section entanglement Chern number for $m/b=1$ (STI phase) along the (a)
$k_1$, (b) $k_2$, and (c) $k_3$ axes, respectively.
(d) The  section entanglement Chern number for $m/b=-0.1$ (trivial phase) along the $k_3$
axis.
}
\label{f:m_1}%-----------------------------------------------
\end{center}
\end{figure}

\begin{figure}[th]
\begin{center}
\begin{tabular}{c}
 \epsfig{file=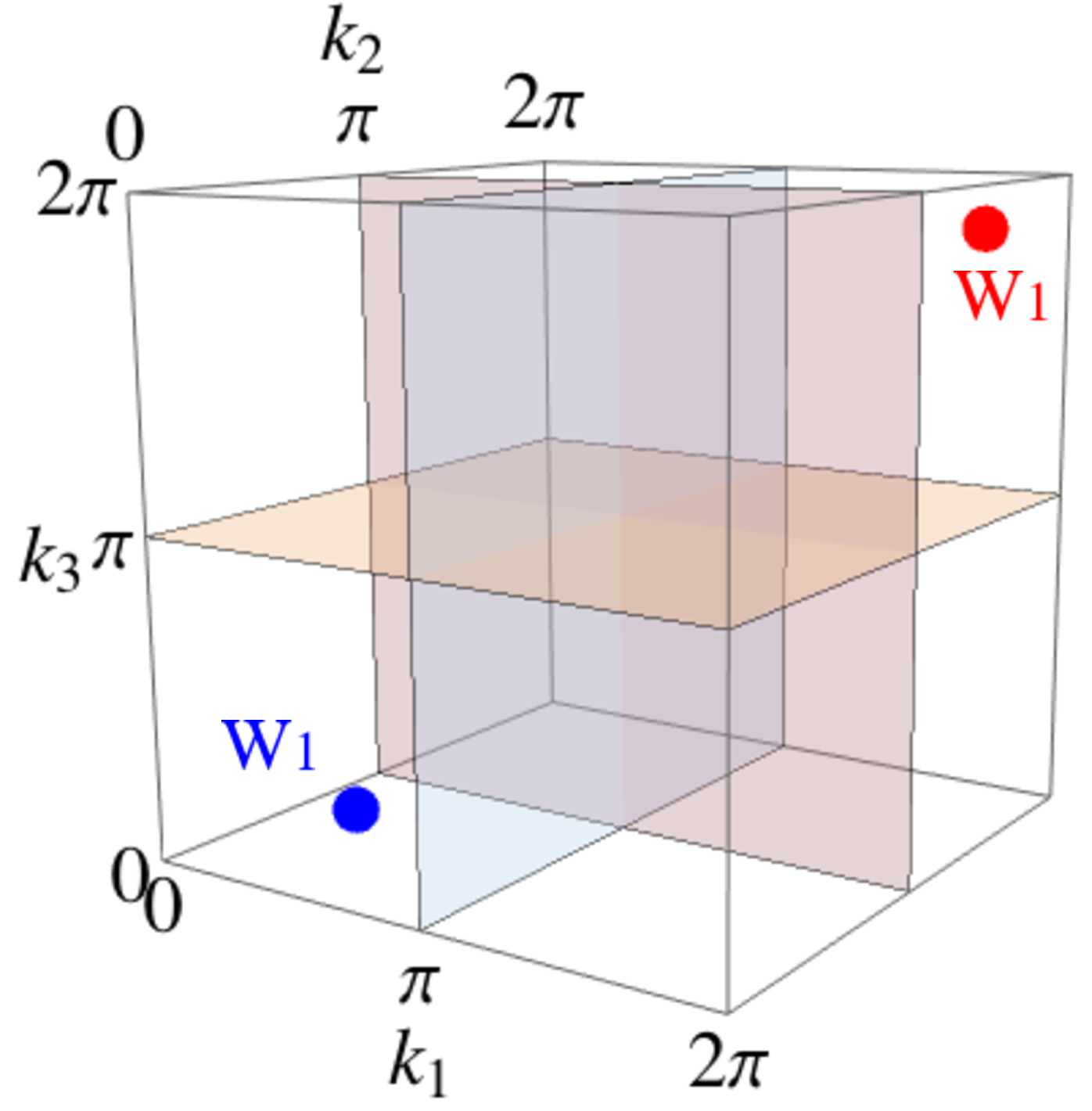,width=0.3\textwidth}
\end{tabular}
 \caption{(Color online)
 Weyl points appearing in Figs. \ref{f:m_1}(a), \ref{f:m_1}(b) and \ref{f:m_1}(c).%\cite{Supp}
}
\label{f:w_m_1}%-----------------------------------------------
\end{center}
\end{figure}

\subsection{STI phase}

We show the section entanglement Chern number in the STI in Figs. \ref{f:m_1}(a),  \ref{f:m_1}(b), and  \ref{f:m_1}(c).
First, from the entanglement Chern number for the time-reversal-invariant planes and the correspondence (\ref{Z2ecCor}),
we see that
$\xi_0=\eta_0=\zeta_0=1$ and $\xi_{\pi}=\eta_{\pi}=\zeta_{\pi}=0$, implying $1(111)$. This is precisely
the known $\mathbb{Z}_2$ number classification given in Table \ref{t:PD}.
Figure \ref{f:m_1} also shows that topological changes occur once between
$0<k_j<\pi$, and once between $\pi<k_j<2\pi$ for $j=1,2,3$.
This implies that in the three-dimensional Brillouin zone, two gapless Weyl points appear in the
entanglement spectrum.
Thus, the topological phase is summarized as Weyl points in Fig. \ref{f:w_m_1}.
The precise position and/or the number of the Weyl points in the whole Brillouin zone depends on
the partition we choose.
However, the number of the Weyl points in the half Brillouin zone, $N_W^{(i)}$, is universal modulo 2.

The Weyl points are also useful to see the topological phase transition.
Let us examine the transition from the STI phase to an ordinary insulating phase at $m/b=0$.
Starting from Fig. \ref{f:m_1}(c), and
moving $m/b\rightarrow0$, the Weyl point
denoted by the red $W_1$ is approaching $k_3\rightarrow2\pi$,
and when $m/b$ switches to a negative value,
the red $W_1$ goes beyond $k_3=2\pi$, as shown in Fig. \ref{f:m_1}(d).
This implies $\zeta_0=\zeta_{\pi}=0$.
Decreasing $m/b$ a bit more, a pair annihilation of $W_1$'s occurs, and section entanglement Chern number becomes trivial all along
the $k_3$ axis. A similar Weyl pair annihilation occurs
along the other $k_1$ and $k_2$ axis.
Thus, the topological change from STI phase to trivial phase can be described by the motion of the Weyl
points in the entanglement spectrum.

\begin{figure}[th]
\begin{center}
 \epsfig{file=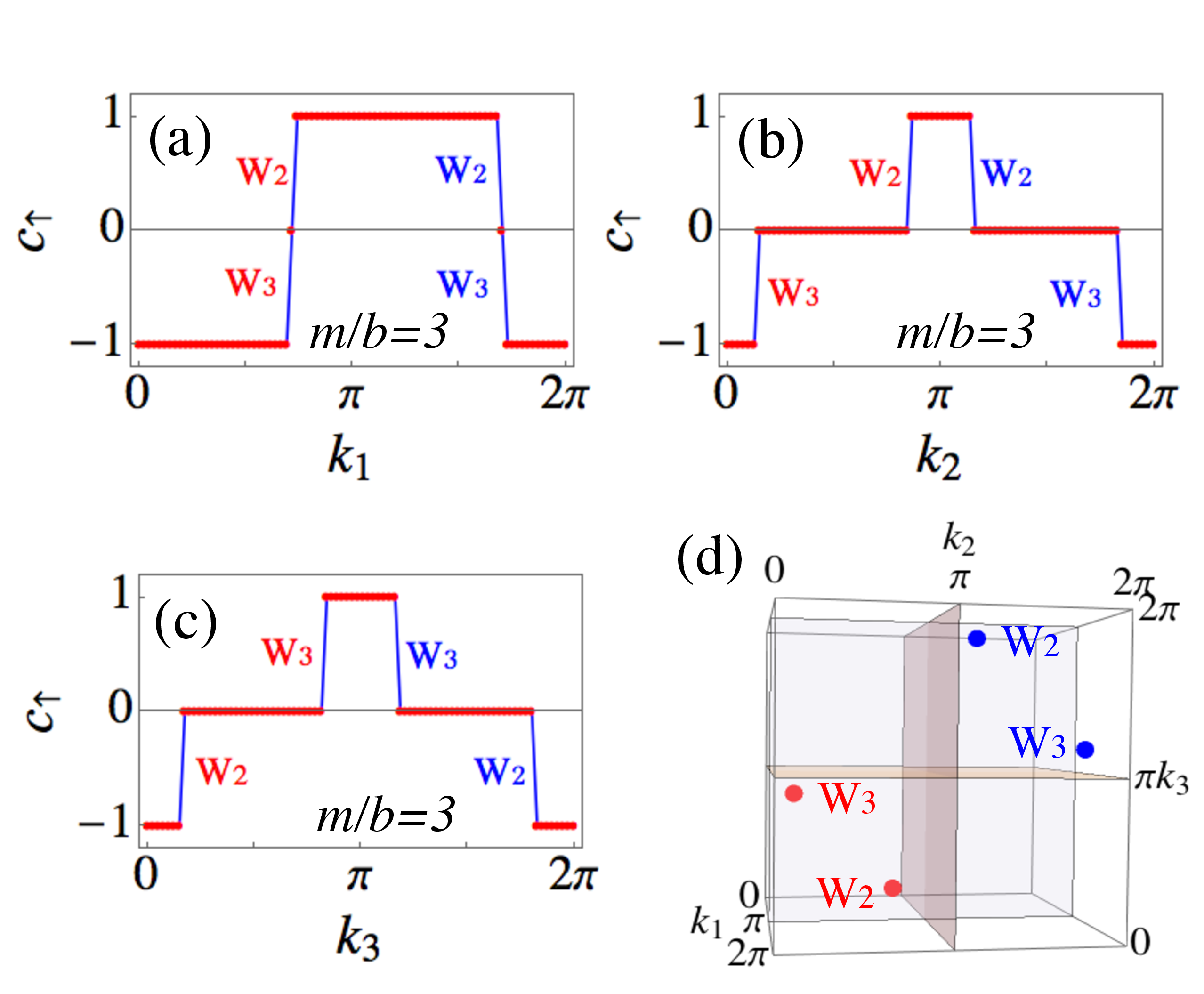,width=0.48\textwidth}
 \caption{(Color online)
 Section entanglement Chern number for $m/b=3$ (WTI phase) along the (a) $k_1$, (b) $k_2$, and  (c) $k_3$
 axes, respectively.
 (d) The Weyl points in (a), (b), and (c).%\cite{Supp}
 }
\label{f:m_3}%-----------------------------------------------
\end{center}
\end{figure}

\subsection{WTI phase}

Next, we analyze the WTI phase.
We show in Figs. \ref{f:m_3}(a),  \ref{f:m_3}(b), and  \ref{f:m_3}(c), 
the section entanglement Chern number  at $m/b=3$.
There appear two Weyl points in $0<k_j<\pi$ and  $\pi<k_j<2\pi$, and thus
the entanglement Chern number for the time-reversal-invariant planes changes by 2.
From the correspondence (\ref{Z2ecCor}), these figures tell us that
$\xi_0=\eta_0=\zeta_0=1$ and $\xi_{\pi}=\eta_{\pi}=\zeta_{\pi}=1$, implying $0(111)$, i.e., WTI phase.
The Weyl points in the entanglement spectrum
are summarized in Fig. \ref{f:m_3}(d). This may be a typical Weyl points in the WTI phase,
and is clearly distinct from Fig. \ref{f:w_m_1}.

\begin{figure}[th]
 \begin{center}
  \epsfig{file=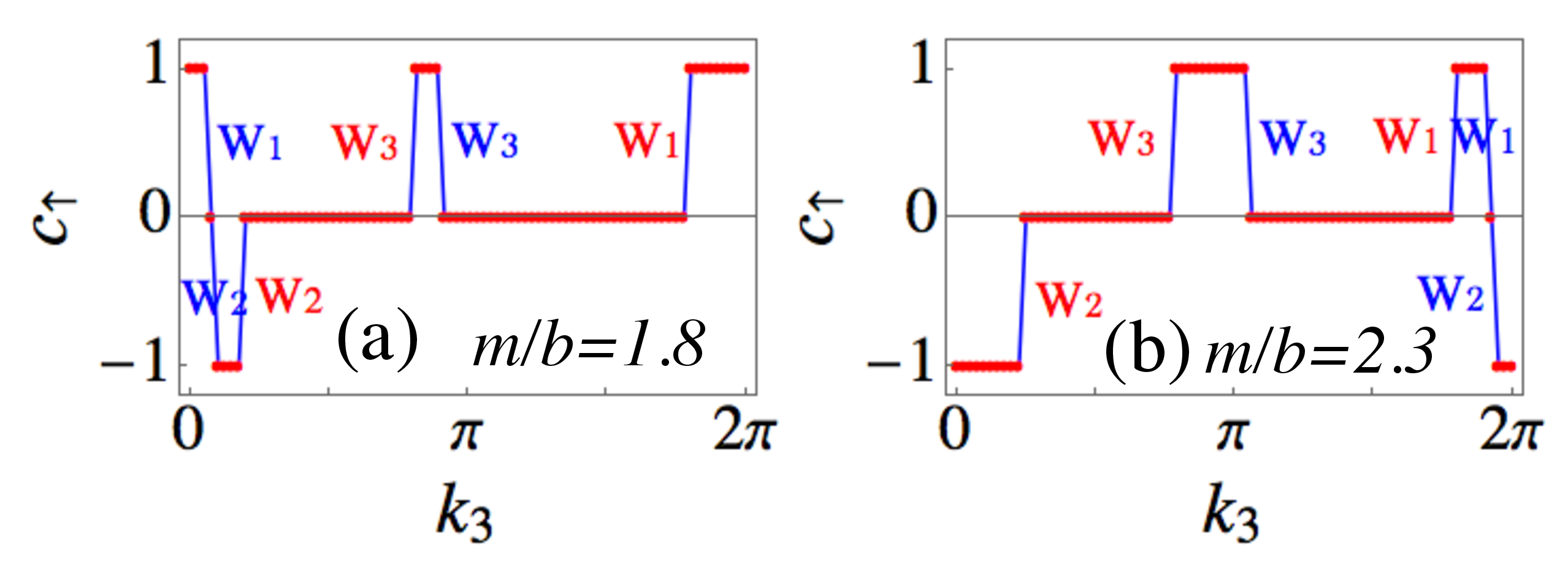,width=0.48\textwidth}
  \caption{(Color online)
  Section entanglement Chern number along the  $k_3$ axis near the phase boundary at $m/b=2$:
  (a) for $m/b=1.8$ still in the STI phase, and  (b) for $m/b=2.3$ just in the WTI phase.
  }
  \label{f:m_3_c}%-----------------------------------------------
 \end{center}
\end{figure}

The topological change between the STI phase and WTI phase is also understood by the behavior
of the Weyl points.
Let us start from the STI phase in Fig. \ref{f:m_1} (c) simply with one pair of Weyl points denoted by
red and blue $W_1$.
Increasing $m/b\rightarrow 2$, pair creations occur in $0<k_3<\pi$,
and as a result, a new pair of Weyl points
denoted by $W_2$ and $W_3$ appears, as shown in Fig. \ref{f:m_3_c}(a).
It should be noted that entanglement Chern numbers at $k_3=0,~\pi$
still show that
$\zeta_0=1$ and $\zeta_{\pi}=0$, implying the model remains in the STI phase.
However,  once $m/b$ goes beyond 2, all blue Weyl points move beyond time-reversal-invariant momenta:
two blue Weyl points,
This causes the change of the topological numbers, $\zeta_0=\zeta_{\pi}=1$, i.e., the WTI phase.
$W_1$ and $W_2$, go beyond $k_3=0$ and one blue $W_3$ goes beyond $k_3=\pi$, as shown in (b).
Increasing $m/b$ a bit more, a pair annihilation of $W_1$ occurs and we reach Fig. \ref{f:m_3}(c).

\section{Conclusion and discussion}
\label{s:cl}%-------

We have investigated the topological insulator phases in three dimensions
in view of the quantum entanglement between extensive up-spin and down-spin electrons.
The Weyl points of the entanglement Hamiltonian characterize the STI and WTI phases.
While the $\mathbb{Z}_2$ invariants are well defined only on the time-reversal-invariant planes, the section entanglement Chern number can be calculated at any section
which interpolates the $\mathbb{Z}_2$ invariants continuously.
This enables us to introduce an idea of Weyl points of the entanglement Hamiltonian.
The emergence of the Weyl points, or the phase transition to the Weyl semimetal, is peculiar to
three dimensions.
For the topological insulator phase, this phenomenon occur for the entanglement Hamiltonians,
which have less symmetry than the
original Hamiltonian.
The STI and WTI have odd and even  $N_W^{(i)}$, respectively.
The section entanglement Chern number changes by $1$ ($-1$), corresponding to the Weyl point of chirality $+$ ($-$)
We have demonstrated these features investigating the Fu--Kane--Mele model 
for the extensive spin partition. We also discuss the generalization of the entanglement Chern number to the pseudospin partition. We have numerically shown the validity of 
the pseudospin partition by testing the Wilson-Dirac model.

We note that a suitable partition should be taken to make an entanglement Hamiltonian. The essential conditions for the entanglement Chern number to work well are that the band inversion of the entanglement Hamiltonian occurs at time-reversal-invariant points and the entanglement spectrum is gapped on the time-reversal-invariant planes. The partition for spin space is quite reasonable from the viewpoint of physical meanings. Actually, the characterization by the entanglement Chern number for spin partition works well for the Fu--Kane--Mele model. For the Wilson-Dirac model, the entanglement spectrum for a (pseudo)spin partition is gapless on a time-reversal-invariant plane. However, the rotation of the spin space gaps out the entanglement spectrum on entire time-reversal-invariant planes, and the characterization by the entanglement Chern number works well for the rotated pseudospin partition. One of the future tasks is to investigate the efficiency of the characterization by the entanglement Chern number for a general model.

One of the advantages of using the entanglement Chern number is that the entanglement Chern number is potentially applicable to topological insulators with interactions.
The topological phases for interacting topological insulators have been well investigated. \cite{
PhysRevB.81.134509,PhysRevLett.105.256803,PhysRevB.83.075103,PhysRevB.83.075102,PhysRevB.86.125119, PhysRevB.93.115131,PhysRevLett.118.147001}
To characterize them by the topological indices, the expansion of the $\mathbb{Z}_2$ topological index for noninteracting topological insulators to that for the interacting topological insulators is nontrivial, since for even-particle systems,  the time-reversal operator becomes $\Theta^2=1$ on the many-body ground state wave functions.
However, the (section) Chern number is well defined even for such interacting systems by using the twisted boundary conditions.\cite{PhysRevB.31.3372} It is thus anticipated that the characterization of topological insulators by the changes of the section entanglement Chern number (the Weyl points of the entanglement Hamiltonian) is useful for some classes of interacting topological insulators, at least for adiabatically deformed systems from the noninteracting topological insulator.\cite{PhysRevLett.100.156401,10.1038/nphys1606,PhysRevLett.104.106408,PhysRevB.85.045130}

Another advantage of using the entanglement Chern number is that the entanglement Chern number is applicable for topological insulators with broken time-reversal-invariant perturbations. We basically believe that topological insulators are stable against very small perturbations even with broken time-reversal symmetry. However, the $\mathbb{Z}_2$ invariance
cannot be defined for such a case, since it is a symmetry-protected topological number. On the other hand, the entanglement Chern number is of course well defined in spite of broken time-reversal symmetry. As we have shown in this paper, the $\mathbb{Z}_2$ invariance and the entanglement Chern number give the same phase diagram for the topological insulators within time-reversal invariance, and the entanglement Chern number can be continued to directions in broken time-reversal invariance. This implies that the entanglement Chern number would give a criterion for the stability of the topological insulators for broken time-reversal perturbations. Indeed, for the two-dimensional case, we have investigated the stability of a topological insulator  against a magnetic field in Ref.~\onlinecite{doi:10.7566/JPSJ.85.043706}. It may be interesting to extend such an analysis to three-dimensional topological insulators with various kinds of perturbations.

%In future, it may be interesting to study the robustness of a topological insulator against
%perturbations with broken time-reversal invariance.
%Further interesting and important problem is to investigate a topological insulator of interacting systems.

\section*{Acknowledgments}
This work is partly supported by Grants-in-Aid for Scientific Research,
No. 17H06138, No. 16K13845 (YH, HA), No. 25107005 (YH, HA), No. 17K05563 (TF), No. 26247064, and No. 25400388 (TF)
from JSPS.

%\nocite{*}


%\bibliographystyle{apsrev4-1}
%\bibliography{ECN}

\begin{thebibliography}{55}%
\makeatletter
\providecommand \@ifxundefined [1]{%
 \@ifx{#1\undefined}
}%
\providecommand \@ifnum [1]{%
 \ifnum #1\expandafter \@firstoftwo
 \else \expandafter \@secondoftwo
 \fi
}%
\providecommand \@ifx [1]{%
 \ifx #1\expandafter \@firstoftwo
 \else \expandafter \@secondoftwo
 \fi
}%
\providecommand \natexlab [1]{#1}%
\providecommand \enquote  [1]{``#1''}%
\providecommand \bibnamefont  [1]{#1}%
\providecommand \bibfnamefont [1]{#1}%
\providecommand \citenamefont [1]{#1}%
\providecommand \href@noop [0]{\@secondoftwo}%
\providecommand \href [0]{\begingroup \@sanitize@url \@href}%
\providecommand \@href[1]{\@@startlink{#1}\@@href}%
\providecommand \@@href[1]{\endgroup#1\@@endlink}%
\providecommand \@sanitize@url [0]{\catcode `\\12\catcode `\$12\catcode
  `\&12\catcode `\#12\catcode `\^12\catcode `\_12\catcode `\%12\relax}%
\providecommand \@@startlink[1]{}%
\providecommand \@@endlink[0]{}%
\providecommand \url  [0]{\begingroup\@sanitize@url \@url }%
\providecommand \@url [1]{\endgroup\@href {#1}{\urlprefix }}%
\providecommand \urlprefix  [0]{URL }%
\providecommand \Eprint [0]{\href }%
\providecommand \doibase [0]{http://dx.doi.org/}%
\providecommand \selectlanguage [0]{\@gobble}%
\providecommand \bibinfo  [0]{\@secondoftwo}%
\providecommand \bibfield  [0]{\@secondoftwo}%
\providecommand \translation [1]{[#1]}%
\providecommand \BibitemOpen [0]{}%
\providecommand \bibitemStop [0]{}%
\providecommand \bibitemNoStop [0]{.\EOS\space}%
\providecommand \EOS [0]{\spacefactor3000\relax}%
\providecommand \BibitemShut  [1]{\csname bibitem#1\endcsname}%
\let\auto@bib@innerbib\@empty
%</preamble>
\bibitem [{\citenamefont {Hasan}\ and\ \citenamefont
  {Kane}(2010)}]{Hasan:2010fk}%
  \BibitemOpen
  \bibfield  {author} {\bibinfo {author} {\bibfnamefont {M.~Z.}\ \bibnamefont
  {Hasan}}\ and\ \bibinfo {author} {\bibfnamefont {C.~L.}\ \bibnamefont
  {Kane}},\ }\href {http://link.aps.org/doi/10.1103/RevModPhys.82.3045}
  {\bibfield  {journal} {\bibinfo  {journal} {Rev. Mod. Phys.}\ }\textbf
  {\bibinfo {volume} {82}},\ \bibinfo {pages} {3045} (\bibinfo {year}
  {2010})}\BibitemShut {NoStop}%
\bibitem [{\citenamefont {Qi}\ and\ \citenamefont {Zhang}(2011)}]{Qi:2011kx}%
  \BibitemOpen
  \bibfield  {author} {\bibinfo {author} {\bibfnamefont {X.-L.}\ \bibnamefont
  {Qi}}\ and\ \bibinfo {author} {\bibfnamefont {S.-C.}\ \bibnamefont {Zhang}},\
  }\href {http://link.aps.org/doi/10.1103/RevModPhys.83.1057} {\bibfield
  {journal} {\bibinfo  {journal} {Rev. Mod. Phys.}\ }\textbf {\bibinfo {volume}
  {83}},\ \bibinfo {pages} {1057} (\bibinfo {year} {2011})}\BibitemShut
  {NoStop}%
\bibitem [{\citenamefont {Schnyder}\ \emph {et~al.}(2008)\citenamefont
  {Schnyder}, \citenamefont {Ryu}, \citenamefont {Furusaki},\ and\
  \citenamefont {Ludwig}}]{PhysRevB.78.195125}%
  \BibitemOpen
  \bibfield  {author} {\bibinfo {author} {\bibfnamefont {A.~P.}\ \bibnamefont
  {Schnyder}}, \bibinfo {author} {\bibfnamefont {S.}~\bibnamefont {Ryu}},
  \bibinfo {author} {\bibfnamefont {A.}~\bibnamefont {Furusaki}}, \ and\
  \bibinfo {author} {\bibfnamefont {A.~W.~W.}\ \bibnamefont {Ludwig}},\ }\href
  {\doibase 10.1103/PhysRevB.78.195125} {\bibfield  {journal} {\bibinfo
  {journal} {Phys. Rev. B}\ }\textbf {\bibinfo {volume} {78}},\ \bibinfo
  {pages} {195125} (\bibinfo {year} {2008})}\BibitemShut {NoStop}%
\bibitem [{\citenamefont {Qi}\ \emph {et~al.}(2008)\citenamefont {Qi},
  \citenamefont {Hughes},\ and\ \citenamefont {Zhang}}]{PhysRevB.78.195424}%
  \BibitemOpen
  \bibfield  {author} {\bibinfo {author} {\bibfnamefont {X.-L.}\ \bibnamefont
  {Qi}}, \bibinfo {author} {\bibfnamefont {T.~L.}\ \bibnamefont {Hughes}}, \
  and\ \bibinfo {author} {\bibfnamefont {S.-C.}\ \bibnamefont {Zhang}},\ }\href
  {\doibase 10.1103/PhysRevB.78.195424} {\bibfield  {journal} {\bibinfo
  {journal} {Phys. Rev. B}\ }\textbf {\bibinfo {volume} {78}},\ \bibinfo
  {pages} {195424} (\bibinfo {year} {2008})}\BibitemShut {NoStop}%
\bibitem [{\citenamefont {Kitaev}(2009)}]{DOI:10.1063/1.3149495}%
  \BibitemOpen
  \bibfield  {author} {\bibinfo {author} {\bibfnamefont {A.}~\bibnamefont
  {Kitaev}},\ }\href {\doibase 10.1063/1.3149495} {\bibfield  {journal}
  {\bibinfo  {journal} {AIP Conf. Proc.}\ }\textbf {\bibinfo {volume} {1134}},\
  \bibinfo {pages} {22} (\bibinfo {year} {2009})}\BibitemShut {NoStop}%
\bibitem [{\citenamefont {Ryu}\ \emph {et~al.}(2010)\citenamefont {Ryu},
  \citenamefont {Schnyder}, \citenamefont {Furusaki},\ and\ \citenamefont
  {Ludwig}}]{1367-2630-12-6-065010}%
  \BibitemOpen
  \bibfield  {author} {\bibinfo {author} {\bibfnamefont {S.}~\bibnamefont
  {Ryu}}, \bibinfo {author} {\bibfnamefont {A.~P.}\ \bibnamefont {Schnyder}},
  \bibinfo {author} {\bibfnamefont {A.}~\bibnamefont {Furusaki}}, \ and\
  \bibinfo {author} {\bibfnamefont {A.~W.~W.}\ \bibnamefont {Ludwig}},\ }\href
  {http://stacks.iop.org/1367-2630/12/i=6/a=065010} {\bibfield  {journal}
  {\bibinfo  {journal} {New J. Phys.}\ }\textbf {\bibinfo {volume} {12}},\
  \bibinfo {pages} {065010} (\bibinfo {year} {2010})}\BibitemShut {NoStop}%
\bibitem [{\citenamefont {Raghu}\ \emph {et~al.}(2008)\citenamefont {Raghu},
  \citenamefont {Qi}, \citenamefont {Honerkamp},\ and\ \citenamefont
  {Zhang}}]{PhysRevLett.100.156401}%
  \BibitemOpen
  \bibfield  {author} {\bibinfo {author} {\bibfnamefont {S.}~\bibnamefont
  {Raghu}}, \bibinfo {author} {\bibfnamefont {X.-L.}\ \bibnamefont {Qi}},
  \bibinfo {author} {\bibfnamefont {C.}~\bibnamefont {Honerkamp}}, \ and\
  \bibinfo {author} {\bibfnamefont {S.-C.}\ \bibnamefont {Zhang}},\ }\href
  {\doibase 10.1103/PhysRevLett.100.156401} {\bibfield  {journal} {\bibinfo
  {journal} {Phys. Rev. Lett.}\ }\textbf {\bibinfo {volume} {100}},\ \bibinfo
  {pages} {156401} (\bibinfo {year} {2008})}\BibitemShut {NoStop}%
\bibitem [{\citenamefont {Pesin}\ and\ \citenamefont
  {Balents}(2010)}]{10.1038/nphys1606}%
  \BibitemOpen
  \bibfield  {author} {\bibinfo {author} {\bibfnamefont {D.}~\bibnamefont
  {Pesin}}\ and\ \bibinfo {author} {\bibfnamefont {L.}~\bibnamefont
  {Balents}},\ }\href {http://dx.doi.org/10.1038/nphys1606} {\bibfield
  {journal} {\bibinfo  {journal} {Nat Phys}\ }\textbf {\bibinfo {volume} {6}},\
  \bibinfo {pages} {376} (\bibinfo {year} {2010})}\BibitemShut {NoStop}%
\bibitem [{\citenamefont {Dzero}\ \emph {et~al.}(2010)\citenamefont {Dzero},
  \citenamefont {Sun}, \citenamefont {Galitski},\ and\ \citenamefont
  {Coleman}}]{PhysRevLett.104.106408}%
  \BibitemOpen
  \bibfield  {author} {\bibinfo {author} {\bibfnamefont {M.}~\bibnamefont
  {Dzero}}, \bibinfo {author} {\bibfnamefont {K.}~\bibnamefont {Sun}}, \bibinfo
  {author} {\bibfnamefont {V.}~\bibnamefont {Galitski}}, \ and\ \bibinfo
  {author} {\bibfnamefont {P.}~\bibnamefont {Coleman}},\ }\href {\doibase
  10.1103/PhysRevLett.104.106408} {\bibfield  {journal} {\bibinfo  {journal}
  {Phys. Rev. Lett.}\ }\textbf {\bibinfo {volume} {104}},\ \bibinfo {pages}
  {106408} (\bibinfo {year} {2010})}\BibitemShut {NoStop}%
\bibitem [{\citenamefont {Fidkowski}\ and\ \citenamefont
  {Kitaev}(2010)}]{PhysRevB.81.134509}%
  \BibitemOpen
  \bibfield  {author} {\bibinfo {author} {\bibfnamefont {L.}~\bibnamefont
  {Fidkowski}}\ and\ \bibinfo {author} {\bibfnamefont {A.}~\bibnamefont
  {Kitaev}},\ }\href {\doibase 10.1103/PhysRevB.81.134509} {\bibfield
  {journal} {\bibinfo  {journal} {Phys. Rev. B}\ }\textbf {\bibinfo {volume}
  {81}},\ \bibinfo {pages} {134509} (\bibinfo {year} {2010})}\BibitemShut
  {NoStop}%
\bibitem [{\citenamefont {Wang}\ \emph {et~al.}(2010)\citenamefont {Wang},
  \citenamefont {Qi},\ and\ \citenamefont {Zhang}}]{PhysRevLett.105.256803}%
  \BibitemOpen
  \bibfield  {author} {\bibinfo {author} {\bibfnamefont {Z.}~\bibnamefont
  {Wang}}, \bibinfo {author} {\bibfnamefont {X.-L.}\ \bibnamefont {Qi}}, \ and\
  \bibinfo {author} {\bibfnamefont {S.-C.}\ \bibnamefont {Zhang}},\ }\href
  {\doibase 10.1103/PhysRevLett.105.256803} {\bibfield  {journal} {\bibinfo
  {journal} {Phys. Rev. Lett.}\ }\textbf {\bibinfo {volume} {105}},\ \bibinfo
  {pages} {256803} (\bibinfo {year} {2010})}\BibitemShut {NoStop}%
\bibitem [{\citenamefont {Fidkowski}\ and\ \citenamefont
  {Kitaev}(2011)}]{PhysRevB.83.075103}%
  \BibitemOpen
  \bibfield  {author} {\bibinfo {author} {\bibfnamefont {L.}~\bibnamefont
  {Fidkowski}}\ and\ \bibinfo {author} {\bibfnamefont {A.}~\bibnamefont
  {Kitaev}},\ }\href {\doibase 10.1103/PhysRevB.83.075103} {\bibfield
  {journal} {\bibinfo  {journal} {Phys. Rev. B}\ }\textbf {\bibinfo {volume}
  {83}},\ \bibinfo {pages} {075103} (\bibinfo {year} {2011})}\BibitemShut
  {NoStop}%
\bibitem [{\citenamefont {Turner}\ \emph {et~al.}(2011)\citenamefont {Turner},
  \citenamefont {Pollmann},\ and\ \citenamefont {Berg}}]{PhysRevB.83.075102}%
  \BibitemOpen
  \bibfield  {author} {\bibinfo {author} {\bibfnamefont {A.~M.}\ \bibnamefont
  {Turner}}, \bibinfo {author} {\bibfnamefont {F.}~\bibnamefont {Pollmann}}, \
  and\ \bibinfo {author} {\bibfnamefont {E.}~\bibnamefont {Berg}},\ }\href
  {\doibase 10.1103/PhysRevB.83.075102} {\bibfield  {journal} {\bibinfo
  {journal} {Phys. Rev. B}\ }\textbf {\bibinfo {volume} {83}},\ \bibinfo
  {pages} {075102} (\bibinfo {year} {2011})}\BibitemShut {NoStop}%
\bibitem [{\citenamefont {Dzero}\ \emph {et~al.}(2012)\citenamefont {Dzero},
  \citenamefont {Sun}, \citenamefont {Coleman},\ and\ \citenamefont
  {Galitski}}]{PhysRevB.85.045130}%
  \BibitemOpen
  \bibfield  {author} {\bibinfo {author} {\bibfnamefont {M.}~\bibnamefont
  {Dzero}}, \bibinfo {author} {\bibfnamefont {K.}~\bibnamefont {Sun}}, \bibinfo
  {author} {\bibfnamefont {P.}~\bibnamefont {Coleman}}, \ and\ \bibinfo
  {author} {\bibfnamefont {V.}~\bibnamefont {Galitski}},\ }\href {\doibase
  10.1103/PhysRevB.85.045130} {\bibfield  {journal} {\bibinfo  {journal} {Phys.
  Rev. B}\ }\textbf {\bibinfo {volume} {85}},\ \bibinfo {pages} {045130}
  (\bibinfo {year} {2012})}\BibitemShut {NoStop}%
\bibitem [{\citenamefont {Lu}\ and\ \citenamefont
  {Vishwanath}(2012)}]{PhysRevB.86.125119}%
  \BibitemOpen
  \bibfield  {author} {\bibinfo {author} {\bibfnamefont {Y.-M.}\ \bibnamefont
  {Lu}}\ and\ \bibinfo {author} {\bibfnamefont {A.}~\bibnamefont
  {Vishwanath}},\ }\href {\doibase 10.1103/PhysRevB.86.125119} {\bibfield
  {journal} {\bibinfo  {journal} {Phys. Rev. B}\ }\textbf {\bibinfo {volume}
  {86}},\ \bibinfo {pages} {125119} (\bibinfo {year} {2012})}\BibitemShut
  {NoStop}%
\bibitem [{\citenamefont {Lapa}\ \emph {et~al.}(2016)\citenamefont {Lapa},
  \citenamefont {Teo},\ and\ \citenamefont {Hughes}}]{PhysRevB.93.115131}%
  \BibitemOpen
  \bibfield  {author} {\bibinfo {author} {\bibfnamefont {M.~F.}\ \bibnamefont
  {Lapa}}, \bibinfo {author} {\bibfnamefont {J.~C.~Y.}\ \bibnamefont {Teo}}, \
  and\ \bibinfo {author} {\bibfnamefont {T.~L.}\ \bibnamefont {Hughes}},\
  }\href {\doibase 10.1103/PhysRevB.93.115131} {\bibfield  {journal} {\bibinfo
  {journal} {Phys. Rev. B}\ }\textbf {\bibinfo {volume} {93}},\ \bibinfo
  {pages} {115131} (\bibinfo {year} {2016})}\BibitemShut {NoStop}%
\bibitem [{\citenamefont {Yoshida}\ \emph {et~al.}(2017)\citenamefont
  {Yoshida}, \citenamefont {Daido}, \citenamefont {Yanase},\ and\ \citenamefont
  {Kawakami}}]{PhysRevLett.118.147001}%
  \BibitemOpen
  \bibfield  {author} {\bibinfo {author} {\bibfnamefont {T.}~\bibnamefont
  {Yoshida}}, \bibinfo {author} {\bibfnamefont {A.}~\bibnamefont {Daido}},
  \bibinfo {author} {\bibfnamefont {Y.}~\bibnamefont {Yanase}}, \ and\ \bibinfo
  {author} {\bibfnamefont {N.}~\bibnamefont {Kawakami}},\ }\href {\doibase
  10.1103/PhysRevLett.118.147001} {\bibfield  {journal} {\bibinfo  {journal}
  {Phys. Rev. Lett.}\ }\textbf {\bibinfo {volume} {118}},\ \bibinfo {pages}
  {147001} (\bibinfo {year} {2017})}\BibitemShut {NoStop}%
\bibitem [{\citenamefont {Hatsugai}(1993)}]{Hatsugai:1993fk}%
  \BibitemOpen
  \bibfield  {author} {\bibinfo {author} {\bibfnamefont {Y.}~\bibnamefont
  {Hatsugai}},\ }\href {http://link.aps.org/doi/10.1103/PhysRevLett.71.3697}
  {\bibfield  {journal} {\bibinfo  {journal} {Phys. Rev. Lett.}\ }\textbf
  {\bibinfo {volume} {71}},\ \bibinfo {pages} {3697} (\bibinfo {year}
  {1993})}\BibitemShut {NoStop}%
\bibitem [{\citenamefont {K{\"o}nig}\ \emph {et~al.}(2007)\citenamefont
  {K{\"o}nig}, \citenamefont {Wiedmann}, \citenamefont {Br{\"u}ne},
  \citenamefont {Roth}, \citenamefont {Buhmann}, \citenamefont {Molenkamp},
  \citenamefont {Qi},\ and\ \citenamefont {Zhang}}]{Konig766}%
  \BibitemOpen
  \bibfield  {author} {\bibinfo {author} {\bibfnamefont {M.}~\bibnamefont
  {K{\"o}nig}}, \bibinfo {author} {\bibfnamefont {S.}~\bibnamefont {Wiedmann}},
  \bibinfo {author} {\bibfnamefont {C.}~\bibnamefont {Br{\"u}ne}}, \bibinfo
  {author} {\bibfnamefont {A.}~\bibnamefont {Roth}}, \bibinfo {author}
  {\bibfnamefont {H.}~\bibnamefont {Buhmann}}, \bibinfo {author} {\bibfnamefont
  {L.~W.}\ \bibnamefont {Molenkamp}}, \bibinfo {author} {\bibfnamefont {X.-L.}\
  \bibnamefont {Qi}}, \ and\ \bibinfo {author} {\bibfnamefont {S.-C.}\
  \bibnamefont {Zhang}},\ }\href {\doibase 10.1126/science.1148047} {\bibfield
  {journal} {\bibinfo  {journal} {Science}\ }\textbf {\bibinfo {volume}
  {318}},\ \bibinfo {pages} {766} (\bibinfo {year} {2007})},\ \Eprint
  {http://arxiv.org/abs/http://science.sciencemag.org/content/318/5851/766.full.pdf}
  {http://science.sciencemag.org/content/318/5851/766.full.pdf} \BibitemShut
  {NoStop}%
\bibitem [{\citenamefont {Steinberg}\ \emph {et~al.}(2011)\citenamefont
  {Steinberg}, \citenamefont {Lalo\"e}, \citenamefont {Fatemi}, \citenamefont
  {Moodera},\ and\ \citenamefont {Jarillo-Herrero}}]{PhysRevB.84.233101}%
  \BibitemOpen
  \bibfield  {author} {\bibinfo {author} {\bibfnamefont {H.}~\bibnamefont
  {Steinberg}}, \bibinfo {author} {\bibfnamefont {J.-B.}\ \bibnamefont
  {Lalo\"e}}, \bibinfo {author} {\bibfnamefont {V.}~\bibnamefont {Fatemi}},
  \bibinfo {author} {\bibfnamefont {J.~S.}\ \bibnamefont {Moodera}}, \ and\
  \bibinfo {author} {\bibfnamefont {P.}~\bibnamefont {Jarillo-Herrero}},\
  }\href {\doibase 10.1103/PhysRevB.84.233101} {\bibfield  {journal} {\bibinfo
  {journal} {Phys. Rev. B}\ }\textbf {\bibinfo {volume} {84}},\ \bibinfo
  {pages} {233101} (\bibinfo {year} {2011})}\BibitemShut {NoStop}%
\bibitem [{\citenamefont {Mourik}\ \emph {et~al.}(2012)\citenamefont {Mourik},
  \citenamefont {Zuo}, \citenamefont {Frolov}, \citenamefont {Plissard},
  \citenamefont {Bakkers},\ and\ \citenamefont {Kouwenhoven}}]{Mourik1003}%
  \BibitemOpen
  \bibfield  {author} {\bibinfo {author} {\bibfnamefont {V.}~\bibnamefont
  {Mourik}}, \bibinfo {author} {\bibfnamefont {K.}~\bibnamefont {Zuo}},
  \bibinfo {author} {\bibfnamefont {S.~M.}\ \bibnamefont {Frolov}}, \bibinfo
  {author} {\bibfnamefont {S.~R.}\ \bibnamefont {Plissard}}, \bibinfo {author}
  {\bibfnamefont {E.~P. A.~M.}\ \bibnamefont {Bakkers}}, \ and\ \bibinfo
  {author} {\bibfnamefont {L.~P.}\ \bibnamefont {Kouwenhoven}},\ }\href
  {\doibase 10.1126/science.1222360} {\bibfield  {journal} {\bibinfo  {journal}
  {Science}\ }\textbf {\bibinfo {volume} {336}},\ \bibinfo {pages} {1003}
  (\bibinfo {year} {2012})},\ \Eprint
  {http://arxiv.org/abs/http://science.sciencemag.org/content/336/6084/1003.full.pdf}
  {http://science.sciencemag.org/content/336/6084/1003.full.pdf} \BibitemShut
  {NoStop}%
\bibitem [{\citenamefont {Fukui}\ \emph {et~al.}(2005)\citenamefont {Fukui},
  \citenamefont {Hatsugai},\ and\ \citenamefont
  {Suzuki}}]{doi:10.1143/JPSJ.74.1674}%
  \BibitemOpen
  \bibfield  {author} {\bibinfo {author} {\bibfnamefont {T.}~\bibnamefont
  {Fukui}}, \bibinfo {author} {\bibfnamefont {Y.}~\bibnamefont {Hatsugai}}, \
  and\ \bibinfo {author} {\bibfnamefont {H.}~\bibnamefont {Suzuki}},\ }\href
  {\doibase 10.1143/JPSJ.74.1674} {\bibfield  {journal} {\bibinfo  {journal}
  {J. Phys. Soc. Jpn.}\ }\textbf {\bibinfo {volume} {74}},\ \bibinfo {pages}
  {1674} (\bibinfo {year} {2005})}\BibitemShut {NoStop}%
\bibitem [{\citenamefont {Wang}\ \emph {et~al.}(2012)\citenamefont {Wang},
  \citenamefont {Qi},\ and\ \citenamefont {Zhang}}]{PhysRevB.85.165126}%
  \BibitemOpen
  \bibfield  {author} {\bibinfo {author} {\bibfnamefont {Z.}~\bibnamefont
  {Wang}}, \bibinfo {author} {\bibfnamefont {X.-L.}\ \bibnamefont {Qi}}, \ and\
  \bibinfo {author} {\bibfnamefont {S.-C.}\ \bibnamefont {Zhang}},\ }\href
  {\doibase 10.1103/PhysRevB.85.165126} {\bibfield  {journal} {\bibinfo
  {journal} {Phys. Rev. B}\ }\textbf {\bibinfo {volume} {85}},\ \bibinfo
  {pages} {165126} (\bibinfo {year} {2012})}\BibitemShut {NoStop}%
\bibitem [{\citenamefont {Wang}\ and\ \citenamefont
  {Zhang}(2012)}]{PhysRevX.2.031008}%
  \BibitemOpen
  \bibfield  {author} {\bibinfo {author} {\bibfnamefont {Z.}~\bibnamefont
  {Wang}}\ and\ \bibinfo {author} {\bibfnamefont {S.-C.}\ \bibnamefont
  {Zhang}},\ }\href {\doibase 10.1103/PhysRevX.2.031008} {\bibfield  {journal}
  {\bibinfo  {journal} {Phys. Rev. X}\ }\textbf {\bibinfo {volume} {2}},\
  \bibinfo {pages} {031008} (\bibinfo {year} {2012})}\BibitemShut {NoStop}%
\bibitem [{\citenamefont {Kane}\ and\ \citenamefont
  {Mele}(2005{\natexlab{a}})}]{PhysRevLett.95.146802}%
  \BibitemOpen
  \bibfield  {author} {\bibinfo {author} {\bibfnamefont {C.~L.}\ \bibnamefont
  {Kane}}\ and\ \bibinfo {author} {\bibfnamefont {E.~J.}\ \bibnamefont
  {Mele}},\ }\href {\doibase 10.1103/PhysRevLett.95.146802} {\bibfield
  {journal} {\bibinfo  {journal} {Phys. Rev. Lett.}\ }\textbf {\bibinfo
  {volume} {95}},\ \bibinfo {pages} {146802} (\bibinfo {year}
  {2005}{\natexlab{a}})}\BibitemShut {NoStop}%
\bibitem [{\citenamefont {Kane}\ and\ \citenamefont
  {Mele}(2005{\natexlab{b}})}]{PhysRevLett.95.226801}%
  \BibitemOpen
  \bibfield  {author} {\bibinfo {author} {\bibfnamefont {C.~L.}\ \bibnamefont
  {Kane}}\ and\ \bibinfo {author} {\bibfnamefont {E.~J.}\ \bibnamefont
  {Mele}},\ }\href {\doibase 10.1103/PhysRevLett.95.226801} {\bibfield
  {journal} {\bibinfo  {journal} {Phys. Rev. Lett.}\ }\textbf {\bibinfo
  {volume} {95}},\ \bibinfo {pages} {226801} (\bibinfo {year}
  {2005}{\natexlab{b}})}\BibitemShut {NoStop}%
\bibitem [{\citenamefont {Fu}\ and\ \citenamefont
  {Kane}(2006)}]{PhysRevB.74.195312}%
  \BibitemOpen
  \bibfield  {author} {\bibinfo {author} {\bibfnamefont {L.}~\bibnamefont
  {Fu}}\ and\ \bibinfo {author} {\bibfnamefont {C.~L.}\ \bibnamefont {Kane}},\
  }\href {\doibase 10.1103/PhysRevB.74.195312} {\bibfield  {journal} {\bibinfo
  {journal} {Phys. Rev. B}\ }\textbf {\bibinfo {volume} {74}},\ \bibinfo
  {pages} {195312} (\bibinfo {year} {2006})}\BibitemShut {NoStop}%
\bibitem [{\citenamefont {Fukui}\ and\ \citenamefont
  {Hatsugai}(2007{\natexlab{a}})}]{Fukui:2007kq}%
  \BibitemOpen
  \bibfield  {author} {\bibinfo {author} {\bibfnamefont {T.}~\bibnamefont
  {Fukui}}\ and\ \bibinfo {author} {\bibfnamefont {Y.}~\bibnamefont
  {Hatsugai}},\ }\href {\doibase 10.1143/JPSJ.76.053702} {\bibfield  {journal}
  {\bibinfo  {journal} {J. Phys. Soc. Jpn.j}\ }\textbf {\bibinfo {volume}
  {76}},\ \bibinfo {pages} {053702} (\bibinfo {year}
  {2007}{\natexlab{a}})}\BibitemShut {NoStop}%
\bibitem [{\citenamefont {Soluyanov}\ and\ \citenamefont
  {Vanderbilt}(2011)}]{Soluyanov:2011aa}%
  \BibitemOpen
  \bibfield  {author} {\bibinfo {author} {\bibfnamefont {A.~A.}\ \bibnamefont
  {Soluyanov}}\ and\ \bibinfo {author} {\bibfnamefont {D.}~\bibnamefont
  {Vanderbilt}},\ }\href {https://link.aps.org/doi/10.1103/PhysRevB.83.235401}
  {\bibfield  {journal} {\bibinfo  {journal} {Phys. Rev. B}\ }\textbf {\bibinfo
  {volume} {83}},\ \bibinfo {pages} {235401} (\bibinfo {year}
  {2011})}\BibitemShut {NoStop}%
\bibitem [{\citenamefont {Vidal}\ \emph {et~al.}(2003)\citenamefont {Vidal},
  \citenamefont {Latorre}, \citenamefont {Rico},\ and\ \citenamefont
  {Kitaev}}]{PhysRevLett.90.227902}%
  \BibitemOpen
  \bibfield  {author} {\bibinfo {author} {\bibfnamefont {G.}~\bibnamefont
  {Vidal}}, \bibinfo {author} {\bibfnamefont {J.~I.}\ \bibnamefont {Latorre}},
  \bibinfo {author} {\bibfnamefont {E.}~\bibnamefont {Rico}}, \ and\ \bibinfo
  {author} {\bibfnamefont {A.}~\bibnamefont {Kitaev}},\ }\href {\doibase
  10.1103/PhysRevLett.90.227902} {\bibfield  {journal} {\bibinfo  {journal}
  {Phys. Rev. Lett.}\ }\textbf {\bibinfo {volume} {90}},\ \bibinfo {pages}
  {227902} (\bibinfo {year} {2003})}\BibitemShut {NoStop}%
\bibitem [{\citenamefont {Ryu}\ and\ \citenamefont
  {Hatsugai}(2006)}]{PhysRevB.73.245115}%
  \BibitemOpen
  \bibfield  {author} {\bibinfo {author} {\bibfnamefont {S.}~\bibnamefont
  {Ryu}}\ and\ \bibinfo {author} {\bibfnamefont {Y.}~\bibnamefont {Hatsugai}},\
  }\href {\doibase 10.1103/PhysRevB.73.245115} {\bibfield  {journal} {\bibinfo
  {journal} {Phys. Rev. B}\ }\textbf {\bibinfo {volume} {73}},\ \bibinfo
  {pages} {245115} (\bibinfo {year} {2006})}\BibitemShut {NoStop}%
\bibitem [{\citenamefont {Kitaev}\ and\ \citenamefont
  {Preskill}(2006)}]{PhysRevLett.96.110404}%
  \BibitemOpen
  \bibfield  {author} {\bibinfo {author} {\bibfnamefont {A.}~\bibnamefont
  {Kitaev}}\ and\ \bibinfo {author} {\bibfnamefont {J.}~\bibnamefont
  {Preskill}},\ }\href {\doibase 10.1103/PhysRevLett.96.110404} {\bibfield
  {journal} {\bibinfo  {journal} {Phys. Rev. Lett.}\ }\textbf {\bibinfo
  {volume} {96}},\ \bibinfo {pages} {110404} (\bibinfo {year}
  {2006})}\BibitemShut {NoStop}%
\bibitem [{\citenamefont {Levin}\ and\ \citenamefont
  {Wen}(2006)}]{PhysRevLett.96.110405}%
  \BibitemOpen
  \bibfield  {author} {\bibinfo {author} {\bibfnamefont {M.}~\bibnamefont
  {Levin}}\ and\ \bibinfo {author} {\bibfnamefont {X.-G.}\ \bibnamefont
  {Wen}},\ }\href {\doibase 10.1103/PhysRevLett.96.110405} {\bibfield
  {journal} {\bibinfo  {journal} {Phys. Rev. Lett.}\ }\textbf {\bibinfo
  {volume} {96}},\ \bibinfo {pages} {110405} (\bibinfo {year}
  {2006})}\BibitemShut {NoStop}%
\bibitem [{\citenamefont {Haque}\ \emph {et~al.}(2007)\citenamefont {Haque},
  \citenamefont {Zozulya},\ and\ \citenamefont
  {Schoutens}}]{PhysRevLett.98.060401}%
  \BibitemOpen
  \bibfield  {author} {\bibinfo {author} {\bibfnamefont {M.}~\bibnamefont
  {Haque}}, \bibinfo {author} {\bibfnamefont {O.}~\bibnamefont {Zozulya}}, \
  and\ \bibinfo {author} {\bibfnamefont {K.}~\bibnamefont {Schoutens}},\ }\href
  {\doibase 10.1103/PhysRevLett.98.060401} {\bibfield  {journal} {\bibinfo
  {journal} {Phys. Rev. Lett.}\ }\textbf {\bibinfo {volume} {98}},\ \bibinfo
  {pages} {060401} (\bibinfo {year} {2007})}\BibitemShut {NoStop}%
\bibitem [{\citenamefont {Zozulya}\ \emph {et~al.}(2007)\citenamefont
  {Zozulya}, \citenamefont {Haque}, \citenamefont {Schoutens},\ and\
  \citenamefont {Rezayi}}]{PhysRevB.76.125310}%
  \BibitemOpen
  \bibfield  {author} {\bibinfo {author} {\bibfnamefont {O.~S.}\ \bibnamefont
  {Zozulya}}, \bibinfo {author} {\bibfnamefont {M.}~\bibnamefont {Haque}},
  \bibinfo {author} {\bibfnamefont {K.}~\bibnamefont {Schoutens}}, \ and\
  \bibinfo {author} {\bibfnamefont {E.~H.}\ \bibnamefont {Rezayi}},\ }\href
  {\doibase 10.1103/PhysRevB.76.125310} {\bibfield  {journal} {\bibinfo
  {journal} {Phys. Rev. B}\ }\textbf {\bibinfo {volume} {76}},\ \bibinfo
  {pages} {125310} (\bibinfo {year} {2007})}\BibitemShut {NoStop}%
\bibitem [{\citenamefont {Friedman}\ and\ \citenamefont
  {Levine}(2008)}]{PhysRevB.78.035320}%
  \BibitemOpen
  \bibfield  {author} {\bibinfo {author} {\bibfnamefont {B.~A.}\ \bibnamefont
  {Friedman}}\ and\ \bibinfo {author} {\bibfnamefont {G.~C.}\ \bibnamefont
  {Levine}},\ }\href {\doibase 10.1103/PhysRevB.78.035320} {\bibfield
  {journal} {\bibinfo  {journal} {Phys. Rev. B}\ }\textbf {\bibinfo {volume}
  {78}},\ \bibinfo {pages} {035320} (\bibinfo {year} {2008})}\BibitemShut
  {NoStop}%
\bibitem [{\citenamefont {Li}\ and\ \citenamefont
  {Haldane}(2008)}]{PhysRevLett.101.010504}%
  \BibitemOpen
  \bibfield  {author} {\bibinfo {author} {\bibfnamefont {H.}~\bibnamefont
  {Li}}\ and\ \bibinfo {author} {\bibfnamefont {F.~D.~M.}\ \bibnamefont
  {Haldane}},\ }\href {\doibase 10.1103/PhysRevLett.101.010504} {\bibfield
  {journal} {\bibinfo  {journal} {Phys. Rev. Lett.}\ }\textbf {\bibinfo
  {volume} {101}},\ \bibinfo {pages} {010504} (\bibinfo {year}
  {2008})}\BibitemShut {NoStop}%
\bibitem [{\citenamefont {Alexandradinata}\ \emph {et~al.}(2011)\citenamefont
  {Alexandradinata}, \citenamefont {Hughes},\ and\ \citenamefont
  {Bernevig}}]{PhysRevB.84.195103}%
  \BibitemOpen
  \bibfield  {author} {\bibinfo {author} {\bibfnamefont {A.}~\bibnamefont
  {Alexandradinata}}, \bibinfo {author} {\bibfnamefont {T.~L.}\ \bibnamefont
  {Hughes}}, \ and\ \bibinfo {author} {\bibfnamefont {B.~A.}\ \bibnamefont
  {Bernevig}},\ }\href {\doibase 10.1103/PhysRevB.84.195103} {\bibfield
  {journal} {\bibinfo  {journal} {Phys. Rev. B}\ }\textbf {\bibinfo {volume}
  {84}},\ \bibinfo {pages} {195103} (\bibinfo {year} {2011})}\BibitemShut
  {NoStop}%
\bibitem [{\citenamefont {Hsieh}\ and\ \citenamefont
  {Fu}(2014)}]{PhysRevLett.113.106801}%
  \BibitemOpen
  \bibfield  {author} {\bibinfo {author} {\bibfnamefont {T.~H.}\ \bibnamefont
  {Hsieh}}\ and\ \bibinfo {author} {\bibfnamefont {L.}~\bibnamefont {Fu}},\
  }\href {\doibase 10.1103/PhysRevLett.113.106801} {\bibfield  {journal}
  {\bibinfo  {journal} {Phys. Rev. Lett.}\ }\textbf {\bibinfo {volume} {113}},\
  \bibinfo {pages} {106801} (\bibinfo {year} {2014})}\BibitemShut {NoStop}%
\bibitem [{\citenamefont {Fukui}\ and\ \citenamefont
  {Hatsugai}(2014)}]{doi:10.7566/JPSJ.83.113705}%
  \BibitemOpen
  \bibfield  {author} {\bibinfo {author} {\bibfnamefont {T.}~\bibnamefont
  {Fukui}}\ and\ \bibinfo {author} {\bibfnamefont {Y.}~\bibnamefont
  {Hatsugai}},\ }\href {\doibase 10.7566/JPSJ.83.113705} {\bibfield  {journal}
  {\bibinfo  {journal} {J. Phys. Soc. Jpn.}\ }\textbf {\bibinfo {volume}
  {83}},\ \bibinfo {pages} {113705} (\bibinfo {year} {2014})}\BibitemShut
  {NoStop}%
\bibitem [{\citenamefont {Haldane}(1988)}]{PhysRevLett.61.2015}%
  \BibitemOpen
  \bibfield  {author} {\bibinfo {author} {\bibfnamefont {F.~D.~M.}\
  \bibnamefont {Haldane}},\ }\href {\doibase 10.1103/PhysRevLett.61.2015}
  {\bibfield  {journal} {\bibinfo  {journal} {Phys. Rev. Lett.}\ }\textbf
  {\bibinfo {volume} {61}},\ \bibinfo {pages} {2015} (\bibinfo {year}
  {1988})}\BibitemShut {NoStop}%
\bibitem [{\citenamefont {Sheng}\ \emph {et~al.}(2006)\citenamefont {Sheng},
  \citenamefont {Weng}, \citenamefont {Sheng},\ and\ \citenamefont
  {Haldane}}]{PhysRevLett.97.036808}%
  \BibitemOpen
  \bibfield  {author} {\bibinfo {author} {\bibfnamefont {D.~N.}\ \bibnamefont
  {Sheng}}, \bibinfo {author} {\bibfnamefont {Z.~Y.}\ \bibnamefont {Weng}},
  \bibinfo {author} {\bibfnamefont {L.}~\bibnamefont {Sheng}}, \ and\ \bibinfo
  {author} {\bibfnamefont {F.~D.~M.}\ \bibnamefont {Haldane}},\ }\href
  {\doibase 10.1103/PhysRevLett.97.036808} {\bibfield  {journal} {\bibinfo
  {journal} {Phys. Rev. Lett.}\ }\textbf {\bibinfo {volume} {97}},\ \bibinfo
  {pages} {036808} (\bibinfo {year} {2006})}\BibitemShut {NoStop}%
\bibitem [{\citenamefont {Fukui}\ and\ \citenamefont
  {Hatsugai}(2007{\natexlab{b}})}]{Fukui:2007sf}%
  \BibitemOpen
  \bibfield  {author} {\bibinfo {author} {\bibfnamefont {T.}~\bibnamefont
  {Fukui}}\ and\ \bibinfo {author} {\bibfnamefont {Y.}~\bibnamefont
  {Hatsugai}},\ }\href {http://link.aps.org/doi/10.1103/PhysRevB.75.121403}
  {\bibfield  {journal} {\bibinfo  {journal} {Phys. Rev. B}\ }\textbf {\bibinfo
  {volume} {75}},\ \bibinfo {pages} {121403} (\bibinfo {year}
  {2007}{\natexlab{b}})}\BibitemShut {NoStop}%
\bibitem [{\citenamefont {Fukui}\ and\ \citenamefont
  {Hatsugai}(2015)}]{doi:10.7566/JPSJ.84.043703}%
  \BibitemOpen
  \bibfield  {author} {\bibinfo {author} {\bibfnamefont {T.}~\bibnamefont
  {Fukui}}\ and\ \bibinfo {author} {\bibfnamefont {Y.}~\bibnamefont
  {Hatsugai}},\ }\href {\doibase 10.7566/JPSJ.84.043703} {\bibfield  {journal}
  {\bibinfo  {journal} {J. Phys. Soc. Jpn.}\ }\textbf {\bibinfo {volume}
  {84}},\ \bibinfo {pages} {043703} (\bibinfo {year} {2015})}\BibitemShut
  {NoStop}%
\bibitem [{\citenamefont {Araki}\ \emph {et~al.}(2016)\citenamefont {Araki},
  \citenamefont {Kariyado}, \citenamefont {Fukui},\ and\ \citenamefont
  {Hatsugai}}]{doi:10.7566/JPSJ.85.043706}%
  \BibitemOpen
  \bibfield  {author} {\bibinfo {author} {\bibfnamefont {H.}~\bibnamefont
  {Araki}}, \bibinfo {author} {\bibfnamefont {T.}~\bibnamefont {Kariyado}},
  \bibinfo {author} {\bibfnamefont {T.}~\bibnamefont {Fukui}}, \ and\ \bibinfo
  {author} {\bibfnamefont {Y.}~\bibnamefont {Hatsugai}},\ }\href {\doibase
  10.7566/JPSJ.85.043706} {\bibfield  {journal} {\bibinfo  {journal} {J. Phys.
  Soc. Jpn.}\ }\textbf {\bibinfo {volume} {85}},\ \bibinfo {pages} {043706}
  (\bibinfo {year} {2016})}\BibitemShut {NoStop}%
\bibitem [{\citenamefont {Young}\ \emph {et~al.}(2012)\citenamefont {Young},
  \citenamefont {Zaheer}, \citenamefont {Teo}, \citenamefont {Kane},
  \citenamefont {Mele},\ and\ \citenamefont {Rappe}}]{PhysRevLett.108.140405}%
  \BibitemOpen
  \bibfield  {author} {\bibinfo {author} {\bibfnamefont {S.~M.}\ \bibnamefont
  {Young}}, \bibinfo {author} {\bibfnamefont {S.}~\bibnamefont {Zaheer}},
  \bibinfo {author} {\bibfnamefont {J.~C.~Y.}\ \bibnamefont {Teo}}, \bibinfo
  {author} {\bibfnamefont {C.~L.}\ \bibnamefont {Kane}}, \bibinfo {author}
  {\bibfnamefont {E.~J.}\ \bibnamefont {Mele}}, \ and\ \bibinfo {author}
  {\bibfnamefont {A.~M.}\ \bibnamefont {Rappe}},\ }\href {\doibase
  10.1103/PhysRevLett.108.140405} {\bibfield  {journal} {\bibinfo  {journal}
  {Phys. Rev. Lett.}\ }\textbf {\bibinfo {volume} {108}},\ \bibinfo {pages}
  {140405} (\bibinfo {year} {2012})}\BibitemShut {NoStop}%
\bibitem [{\citenamefont {Ojanen}(2013)}]{PhysRevB.87.245112}%
  \BibitemOpen
  \bibfield  {author} {\bibinfo {author} {\bibfnamefont {T.}~\bibnamefont
  {Ojanen}},\ }\href {\doibase 10.1103/PhysRevB.87.245112} {\bibfield
  {journal} {\bibinfo  {journal} {Phys. Rev. B}\ }\textbf {\bibinfo {volume}
  {87}},\ \bibinfo {pages} {245112} (\bibinfo {year} {2013})}\BibitemShut
  {NoStop}%
\bibitem [{\citenamefont {Fu}\ \emph {et~al.}(2007)\citenamefont {Fu},
  \citenamefont {Kane},\ and\ \citenamefont {Mele}}]{PhysRevLett.98.106803}%
  \BibitemOpen
  \bibfield  {author} {\bibinfo {author} {\bibfnamefont {L.}~\bibnamefont
  {Fu}}, \bibinfo {author} {\bibfnamefont {C.~L.}\ \bibnamefont {Kane}}, \ and\
  \bibinfo {author} {\bibfnamefont {E.~J.}\ \bibnamefont {Mele}},\ }\href
  {\doibase 10.1103/PhysRevLett.98.106803} {\bibfield  {journal} {\bibinfo
  {journal} {Phys. Rev. Lett.}\ }\textbf {\bibinfo {volume} {98}},\ \bibinfo
  {pages} {106803} (\bibinfo {year} {2007})}\BibitemShut {NoStop}%
\bibitem [{\citenamefont {Roy}(2009)}]{PhysRevB.79.195322}%
  \BibitemOpen
  \bibfield  {author} {\bibinfo {author} {\bibfnamefont {R.}~\bibnamefont
  {Roy}},\ }\href {\doibase 10.1103/PhysRevB.79.195322} {\bibfield  {journal}
  {\bibinfo  {journal} {Phys. Rev. B}\ }\textbf {\bibinfo {volume} {79}},\
  \bibinfo {pages} {195322} (\bibinfo {year} {2009})}\BibitemShut {NoStop}%
\bibitem [{\citenamefont {Moore}\ and\ \citenamefont
  {Balents}(2007)}]{PhysRevB.75.121306}%
  \BibitemOpen
  \bibfield  {author} {\bibinfo {author} {\bibfnamefont {J.~E.}\ \bibnamefont
  {Moore}}\ and\ \bibinfo {author} {\bibfnamefont {L.}~\bibnamefont
  {Balents}},\ }\href {\doibase 10.1103/PhysRevB.75.121306} {\bibfield
  {journal} {\bibinfo  {journal} {Phys. Rev. B}\ }\textbf {\bibinfo {volume}
  {75}},\ \bibinfo {pages} {121306} (\bibinfo {year} {2007})}\BibitemShut
  {NoStop}%
\bibitem [{\citenamefont {Peschel}(2003)}]{0305-4470-36-14-101}%
  \BibitemOpen
  \bibfield  {author} {\bibinfo {author} {\bibfnamefont {I.}~\bibnamefont
  {Peschel}},\ }\href {http://stacks.iop.org/0305-4470/36/i=14/a=101}
  {\bibfield  {journal} {\bibinfo  {journal} {J. Phys. A: Math. Gen.}\ }\textbf
  {\bibinfo {volume} {36}},\ \bibinfo {pages} {L205} (\bibinfo {year}
  {2003})}\BibitemShut {NoStop}%
\bibitem [{\citenamefont {Murakami}(2007)}]{Murakami:2007uq}%
  \BibitemOpen
  \bibfield  {author} {\bibinfo {author} {\bibfnamefont {S.}~\bibnamefont
  {Murakami}},\ }\href {http://arXiv.org/abs/0710.0930} {\bibfield  {journal}
  {\bibinfo  {journal} {New J. Phys.}\ }\textbf {\bibinfo {volume} {9}},\
  \bibinfo {pages} {356} (\bibinfo {year} {2007})},\ \Eprint
  {http://arxiv.org/abs/0710.0930} {0710.0930} \BibitemShut {NoStop}%
\bibitem [{\citenamefont {Burkov}\ and\ \citenamefont
  {Balents}(2011)}]{Burkov:2011ab}%
  \BibitemOpen
  \bibfield  {author} {\bibinfo {author} {\bibfnamefont {A.~A.}\ \bibnamefont
  {Burkov}}\ and\ \bibinfo {author} {\bibfnamefont {L.}~\bibnamefont
  {Balents}},\ }\href {http://link.aps.org/doi/10.1103/PhysRevLett.107.127205}
  {\bibfield  {journal} {\bibinfo  {journal} {Phys. Rev. Lett.}\ }\textbf
  {\bibinfo {volume} {107}},\ \bibinfo {pages} {127205} (\bibinfo {year}
  {2011})}\BibitemShut {NoStop}%
\bibitem [{\citenamefont {Fu}\ and\ \citenamefont
  {Kane}(2007)}]{PhysRevB.76.045302}%
  \BibitemOpen
  \bibfield  {author} {\bibinfo {author} {\bibfnamefont {L.}~\bibnamefont
  {Fu}}\ and\ \bibinfo {author} {\bibfnamefont {C.~L.}\ \bibnamefont {Kane}},\
  }\href {\doibase 10.1103/PhysRevB.76.045302} {\bibfield  {journal} {\bibinfo
  {journal} {Phys. Rev. B}\ }\textbf {\bibinfo {volume} {76}},\ \bibinfo
  {pages} {045302} (\bibinfo {year} {2007})}\BibitemShut {NoStop}%
\bibitem [{\citenamefont {Niu}\ \emph {et~al.}(1985)\citenamefont {Niu},
  \citenamefont {Thouless},\ and\ \citenamefont {Wu}}]{PhysRevB.31.3372}%
  \BibitemOpen
  \bibfield  {author} {\bibinfo {author} {\bibfnamefont {Q.}~\bibnamefont
  {Niu}}, \bibinfo {author} {\bibfnamefont {D.~J.}\ \bibnamefont {Thouless}}, \
  and\ \bibinfo {author} {\bibfnamefont {Y.-S.}\ \bibnamefont {Wu}},\ }\href
  {\doibase 10.1103/PhysRevB.31.3372} {\bibfield  {journal} {\bibinfo
  {journal} {Phys. Rev. B}\ }\textbf {\bibinfo {volume} {31}},\ \bibinfo
  {pages} {3372} (\bibinfo {year} {1985})}\BibitemShut {NoStop}%
\end{thebibliography}

\end{document}